\documentclass{optica-article}

\journal{opticajournal} 

\articletype{Research Article}

\usepackage{lineno}
\usepackage{xcolor}

\begin{document}

\title{Light propagation through an atomic vapor with non-orthogonal electric field modes}

\author{Jack D Briscoe, Danielle Pizzey\authormark{*}, Robert M Potvliege, Steven A Wrathmall, and Ifan G 
 Hughes}

\address{Department of Physics, Durham University, South Road, Durham, DH1 3LE, United Kingdom.}

\email{\authormark{*}danielle.boddy@durham.ac.uk}

\begin{abstract*} 
Alkali-metal atomic vapors are the foundation of an ever-growing range of applications, driven by a comprehensive understanding of their interaction with light. In particular, many models have been developed which characterize this interaction for low intensity laser fields. An atomic medium subject to an external magnetic field of arbitrary direction exhibits two electric field modes that, in general, are non-orthogonal. Mode non-orthogonality is currently neglected by the models used in this context. We derive a new light propagation formalism which takes into account the non-zero overlap of the two modes. We verify the theory using weak-probe spectroscopy of the Rb D$_{2}$ line, showing excellent agreement with experiment. The predictions of the new theory can be exploited, and optimized, to design better atomic photonic devices.
\end{abstract*}

\section{Introduction}
\label{sec: introduction}

The range of applications of alkali-metal atomic vapors is vast, not limited to: electromagnetic field sensing~\cite{fan2015atom, allinson2023simultaneous, jau2020vapor}, frequency conversion for THz imaging~\cite{downes2020full}, atomic clocks~\cite{knappe2006microfabricated, gharavipour2016high}, quantum memories~\cite{lvovsky2009optical, mottola2023optical}, laser stabilization~\cite{chang2022frequency, shi2022frequency}, magnetometry~\cite{griffith2010femtotesla, sutter2020recording, staerkind2024high, auzinsh2022wide} and bandpass filtering~\cite{dick1991ultrahigh, yeh1982dispersive, uhland2023build}. It is therefore crucial that we understand their interactions with light. Many models exist which aim to characterize these interactions in both strong~\cite{downes2023simple, potvliege2025coombe, ADMweb, bala2022comprehensive, sagle1996measurement} and weak-probe~\cite{zentile2015elecsus, keaveney2018elecsus} laser fields. These models encapsulate the response of an atomic medium via the electric susceptibility $\chi$~\cite{adams2018optics}; for weak laser 
fields, the response is analytic and therefore well understood~\cite{siddons2008absolute}. It is therefore expected to find excellent agreement between theory and experiment~\cite{siddons2008absolute, zentile2015atomic, zentile2015optimization, zentile2014hyperfine, hanley2015absolute, ponciano2020absorption, briscoe2023voigt, briscoe2024indirect, pizzey2022laser, uhland2023build, reed2018low, luo2018signal, xiong2018characteristics, agnew2024practical}. 

In most applications, the magnitude of an external magnetic field is integrated into a theory model via a Zeeman shift of atomic resonances~\cite{foot2005atomic}. The magnetic field direction, however, is less commonly modeled, and is often fixed in well known configurations where analytic equations for output intensities are known~\cite{weller2012measuring}. For example, a laser beam propagating parallel to an external magnetic field---used to exploit the Faraday effect~\cite{faraday1846experimental, budker2002resonant}---is often the go-to orientation in many applications~\cite{siddons2009gigahertz, aplet1964faraday, wu1986optical, portalupi2016simultaneous, zentile2015elecsus}. We refer to this orientation as the Faraday geometry. While less common in the literature, the Voigt geometry, where an external magnetic field is perpendicular to the laser beam~\cite{mottola2023electromagnetically, mottola2023optical}, has been explored in the context of spectroscopy~\cite{ponciano2020absorption, briscoe2023voigt, briscoe2024indirect, muroo1994resonant}, atomic filtering~\cite{kudenov2020dual}, and designing frequency-selective lasers~\cite{ge2024voigt, liu2023atomic}. Studies that utilize arbitrary atom-light geometries, defined by the relative angle $\theta_{B}$ between the light wavevector $\mathbf{k}$ and the external magnetic field $\mathbf{B}$, are least common due to the increased complexity of modeling an atom-light system with a non-trivial interaction between the light polarization plane and angular momenta quantization axes~\cite{palik1970infrared, rotondaro2015generalized, keaveney2018elecsus, edwards1995magneto, nienhuis1998magneto}. Recent studies have exploited arbitrary geometries, specifically when $\theta_{B} \approx 80^{\circ}$, to realize atomic bandpass filters with significantly improved performances~\cite{keaveney2018optimized, higgins2020atomic, alqarni2024device}. These studies push the boundary of what is typically capable with a thermally-broadened atomic vapor~\cite{zentile2015atomic}. It is therefore paramount that the same standard of theory-to-experiment agreement is attained for all magnetic field directions, as this allows for parameter optimization of outputs away from the laboratory setting~\cite{zentile2015atomic, zentile2015optimization, keaveney2018optimized, briscoe2023voigt}.

To propagate light through an atomic vapor subject to an external magnetic field of arbitrary direction, a non-trivial wave equation must be solved for both the refractive indices and electric field modes of the medium~\cite{palik1970infrared, rotondaro2015generalized, keaveney2018elecsus}. The modes of such a system are in general both non-orthogonal and frequency dependent~\cite{logueThesis}; the only exceptions are the well-known Faraday and Voigt geometries, whose modes are orthogonal~\cite{briscoe2024indirect}. This non-orthogonality is currently not accounted for by models often used in this context~\cite{rotondaro2015generalized, keaveney2018elecsus}. In this study, we therefore propose a new light propagation formalism to account for non-orthogonal electric field modes, constructed using an electric field matching condition where light enters the atomic vapor. We integrate the new formalism into our model \emph{ElecSus}~\cite{zentile2015elecsus, keaveney2018elecsus}, and experimentally verify it using weak-probe spectroscopy of the Rb D$_{2}$ line. Excellent agreement between the new model and experiment is found. Not only does the new light propagation formalism remove large unphysical features in regions of the strongest mode overlap, we also demonstrate regimes where the old formalism is markedly incorrect across large frequency ranges. This is more notable at intermediate magnetic field geometries between Faraday and Voigt. We therefore show that the non-orthogonality of electric field modes inside the atomic vapor must be taken into account in any valid model of such systems. 

In the remainder of this paper, outputs generated using our new light propagation formalism are labeled ``\emph{ElecSus}~$4$'', while outputs predicted by the existing formalism, in which the non-orthogonality of the modes is neglected, are labeled ``\emph{ElecSus}~$3$''~\cite{keaveney2018elecsus}. These terms refer to the respective versions of the \emph{ElecSus} program~\cite{ElecSusGitHub, zentile2015elecsus, keaveney2018elecsus}, which is often used in weak-probe transmission spectroscopy. The word $\emph{ElecSus}$ refers to the entire model---into which the associated light propagation formalism is implemented.

\section{Theory}
\label{sec: theory}
\subsection{Calculating Refractive Indices and Electric Field Modes}

\begin{figure}
\centering
{\includegraphics[width=\linewidth]{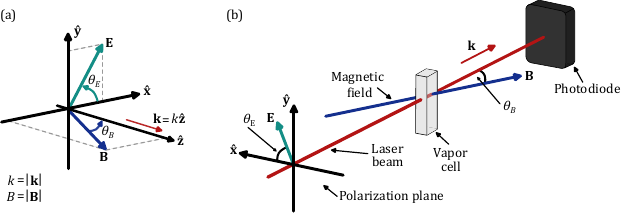}}
\caption{The atom-light system. (a) The coordinate system. Light propagation is calculated in the right-handed laboratory frame $\hat{\mathbf{x}}\hat{\mathbf{y}}\hat{\mathbf{z}}$ in which the $z$-axis is in the direction of the wavevector $\mathbf{k}$ and the external magnetic field $\mathbf{B}$ lies in the $\hat{\mathbf{x}}-\hat{\mathbf{z}}$ plane with relative angle $\theta_{B}$ with respect to $\mathbf{k}$. (b) The coordinate system applied to transmission spectroscopy of an atomic vapor cell immersed in an external magnetic field. A laser beam traverses the vapor cell and propagates in the direction defined by the wavevector $\mathbf{k}$. In this paper, the laser is initially linearly polarized in the direction characterized by the angle $\theta_{E}$ relative to $\hat{\mathbf{x}}$. The transmitted intensity is measured by a photodiode.}
\label{fig: atomicSystem}
\end{figure}

The first part of the theory details a derivation of the refractive indices and electric field modes of an alkali-metal atomic medium subject to an external magnetic field of arbitrary direction, following the literature~\cite{rotondaro2015generalized, keaveney2018elecsus, palik1970infrared}. We model an atomic vapor as a non-magnetic dielectric medium subject to a weak-probe monochromatic plane-wave electric field of the form

\begin{equation}
\mathbf{E}(\mathbf{r}, t) = \mathrm{Re}\{\mathcal{E}_{0}\hat{\boldsymbol {\epsilon}}\,\mathrm{exp}[\mathrm{i}(\mathbf{k}\cdot\mathbf{r} - \omega t)]\} \,,
\end{equation}

\noindent where $\mathcal{E}_{0}$ is the electric field amplitude, $\mathbf{k}$ is the wavevector ($k = \lvert\mathbf{k}\rvert$), $\mathbf{r}$ is a position vector ($r = \lvert\mathbf{r}\rvert$), $\omega$ is the angular frequency of the wave, $t$ is time, and $\mathrm{Re}\{\cdot\}$ enforces the electric field $\mathbf{E}$ in complex notation to be a real quantity. We denote the normalized electric field polarization vector $\hat{\boldsymbol {\epsilon}}$, which is complex, assumed to be independent of $t$, and defined in a general Cartesian coordinate frame labeled $\hat{\mathbf{x}}\hat{\mathbf{y}}\hat{\mathbf{z}}$ as

\begin{equation}
\hat{\boldsymbol {\epsilon}} = \epsilon_{x}\hat{\mathbf{x}} + \epsilon_{y}\hat{\mathbf{y}} + \epsilon_{z}\hat{\mathbf{z}} 
= [\epsilon_{x}, \epsilon_{y}, \epsilon_{z}]^{\mathrm{T}} \,,
\end{equation}

\noindent where $\epsilon_{j}$ are components of the polarization along the direction $j$. The atom-light interaction inside the medium is therefore described by the wave equation~\cite{rotondaro2015generalized} 

\begin{equation} 
\label{eqn: waveEqn}
\mathbf{k} \times (\mathbf{k} \times \mathbf{E}) + \frac{\omega^{2}}{c^{2}}\underline{\varepsilon}\cdot\mathbf{E} = 0 \,,
\end{equation}

\noindent where $c$ is the speed of light in a vacuum, and $\underline{\varepsilon}$ is the dielectric tensor. For simplicity, we take the $z$-axis of the system of coordinates to be in the direction of propagation of the light, so that 
$\mathbf{k} = k\hat{\mathbf{z}}$. We refer to this coordinate system as the laboratory (lab) frame. As shown by the schematic in Fig.~\ref{fig: atomicSystem}(a), this system of axes is right-handed and the external magnetic field $\mathbf{B}$ is in the $\hat{\mathbf{x}}$-$\hat{\mathbf{z}}$ plane, making an angle $\theta_{B}$ with the wavevector $\mathbf{k}$. We note, however, that the elements of the dielectric tensor are more easily calculated in a right-handed Cartesian coordinate frame $\hat{\mathbf{x}}^{\prime}\hat{\mathbf{y}}^{\prime}\hat{\mathbf{z}}^{\prime}$ rotated by $\theta_{B}$ about the $y$-axis with respect to the $\hat{\mathbf{x}}\hat{\mathbf{y}}\hat{\mathbf{z}}$ frame. The $z$-axis of this other system is indeed in the direction of the external magnetic field (i.~e.~$\mathbf{B} = B\hat{\mathbf{z}}^{\prime}$ with $B = \lvert\mathbf{B}\rvert$) and is therefore the natural choice for the angular momenta quantization axis. 

As shown in Ref.~\cite{rotondaro2015generalized}, Eq.~(\ref{eqn: waveEqn}) can be recast in the following form:

\begin{equation} \label{eqn: waveEqnMatrix}
\begin{pmatrix} (\varepsilon^{\prime}_{xx} - n^{2})\,\mathrm{cos}\theta_{B} & \varepsilon^{\prime}_{xy} & \varepsilon^{\prime}_{xx}\,\mathrm{sin}\theta_{B} \\ -\varepsilon^{\prime}_{xy}\,\mathrm{cos}\theta_{B} & \varepsilon^{\prime}_{xx} - n^{2} & -\varepsilon^{\prime}_{xy}\,\mathrm{sin}\theta_{B} \\ -(\varepsilon^{\prime}_{zz} - n^{2})\,\mathrm{sin}\theta_{B} & 0 & \varepsilon^{\prime}_{zz}\,\mathrm{cos}\theta_{B} \end{pmatrix}\begin{pmatrix}
\epsilon_{x} \\
\epsilon_{y} \\
\epsilon_{z}
\end{pmatrix} = 0 \,,
\end{equation}

\noindent where $\varepsilon^{\prime}_{xx}$, $\varepsilon^{\prime}_{xy}$ and $\varepsilon^{\prime}_{zz}$ are components of the dielectric tensor in the $\hat{\mathbf{x}}^{\prime}\hat{\mathbf{y}}^{\prime}\hat{\mathbf{z}}^{\prime}$ system, $n^{2} = (c/\omega)^{2}\mathbf{k}\cdot\mathbf{k} = (1/k_{0}^{2})\mathbf{k}\cdot\mathbf{k}$ is the square of the complex refractive index, and $k_{0} = 2\pi/\lambda_{0}$ is the vacuum wavenumber for light of wavelength $\lambda_{0}$. Here~\cite{rotondaro2015generalized}

\begin{subequations}
\label{eqn: dielectricTensorComponents}
\begin{align}
\varepsilon^{\prime}_{xx} & = 1 + \frac{\chi_{\mathrm{+1}}}{2} + \frac{\chi_{\mathrm{-1}}}{2} \,, \\
\varepsilon^{\prime}_{xy} & = \frac{\mathrm{i}\chi_{\mathrm{-1}}}{2} - \frac{\mathrm{i}\chi_{\mathrm{+1}}}{2} \,, \\
\varepsilon^{\prime}_{zz} & = 1 + \chi_{\mathrm{0}} \,,
\end{align}
\end{subequations}

\noindent where $\chi_{\mathrm{+1}}$, $\chi_{\mathrm{-1}}$ and $\chi_{\mathrm{0}}$ are the complex electric susceptibilities corresponding, respectively, to $\sigma^+$, $\sigma^-$ and $\pi$~transitions \cite{foot2005atomic}. We calculate these quantities as described in ~\cite{zentile2015elecsus, keaveney2018elecsus, wellerThesis}.

The complex refractive indices $n_{i}$ and associated electric field modes $\hat{\mathbf{m}}_{i}$ inside the atomic medium by solving Eq.~(\ref{eqn: waveEqnMatrix}). To this effect, we set the determinant of the matrix to zero, leading to two solutions, $n_{1}$ and $n_{2}$. Each of these two values corresponds to a normalized electric field mode $\hat{\mathbf{m}}_{i} = [m_{i, x}, m_{i, y}, m_{i, z}]^{\mathrm{T}}$, found by solving the wave equation separately for $n_{1}$ and $n_{2}$, i.~e.~

\begin{equation}
\label{eqn: DispersionEquationGetModes}
\begin{pmatrix}
(\varepsilon^{\prime}_{xx} - n^{2}_{i})\,\mathrm{cos}\,\theta_{B} & \varepsilon^{\prime}_{xy} & \varepsilon^{\prime}_{xx}\,\mathrm{sin}\,\theta_{B}\\
-\varepsilon^{\prime}_{xy}\,\mathrm{cos}\,\theta_{B} & \varepsilon^{\prime}_{xx} - n^{2}_{i} & -\varepsilon^{\prime}_{xy}\,\mathrm{sin}\,\theta_{B}\\
-(\varepsilon^{\prime}_{zz} - n^{2}_{i})\,\mathrm{sin}\,\theta_{B} & 0 & \varepsilon^{\prime}_{zz}\,\mathrm{cos}\,\theta_{B}\\
\end{pmatrix}
\begin{pmatrix}
m_{i, x} \\
m_{i, y} \\
m_{i, z} \\
\end{pmatrix}
= 0 \,.
\end{equation}

\noindent Both $\hat{\mathbf{m}}_{1}$ and $\hat{\mathbf{m}}_{2}$---which correspond to the refractive indices $n_1$ and $n_2$ respectively---are therefore particular solutions of the wave equation with a well defined frequency and polarization.
Most importantly, $\hat{\mathbf{m}}_{1}$ and $\hat{\mathbf{m}}_{2}$ form a set of polarization basis vectors that describe how the light propagates through the atomic vapor characterized by the wave equation. Since these modes are solved independently for two different complex refractive indices $n_{i}$, in general they are non-orthogonal (which we explicitly demonstrate in Section~\ref{sec: results}). In the same section, we show that mode non-orthogonality is currently not properly accounted for by the (old) light propagation formalism currently used in this context~\cite{rotondaro2015generalized, keaveney2018elecsus}. We therefore present a new light propagation formalism which does not suffer from this oversight in the next section. Note, the modes $\hat{\mathbf{m}}_{1}$ and $\hat{\mathbf{m}}_{2}$ are always orthogonal in the Faraday geometry ($\theta_{B}=0$) and in the Voigt geometry ($\theta_{B}=\pi/2$)~\cite{briscoe2024indirect}. It can be shown that they would also be orthogonal for any value of $\theta_{B}$ if the susceptibilities $\chi_{+1}$, $\chi_{-1}$ and $\chi_{0}$ were real rather than complex.

\subsection{Light Propagation}
The electric field modes act as a unique polarization basis for the atom-light system, each mode corresponding to a unique electric field $\mathbf{E}_{\hat{\mathbf {m}}_{i}}(z, t)$ that solves the wave equation:

\begin{equation}
\label{eqn: modeEfield}
\mathbf{E}_{\hat{\mathbf {m}}_{i}}(z, t) = \mathrm{Re}\{\mathcal{E}_{0}\hat{\mathbf {m}}_{i}\,\mathrm{exp}[\mathrm{i}(kz - \omega t)]\} \,.
\end{equation}

\noindent Given that $\mathbf{k}\cdot\mathbf{r} = 
kz$ in our choice of coordinate system, we now express the fields as functions of $z$ rather than functions of $\mathbf{r}$. On entering the atomic vapor ($z = 0$), the principle of linear superposition can be used to reconstruct any incident electric field $\mathbf{E}_{\mathrm{inc}}(z,t)$ as a linear combination of $\hat{\mathbf{m}}_{1}$ and $\hat{\mathbf{m}}_{2}$:

\begin{equation}
\label{eqn: incidentE}
\mathbf{E}_{\mathrm{inc}}(z = 0) = \mathcal{E}_{0}\hat{\boldsymbol{\epsilon}}_{\mathrm{inc}} = \mathcal{E}_{0}(c_{1}\hat{\mathbf{m}}_{1} + c_{2}\hat{\mathbf{m}}_{2}) \,,
\end{equation}

\noindent where $c_{i}$ are complex coefficients, $\hat{\mathbf{m}}_{i}$ are assumed to be normalized, $\hat{\boldsymbol{\epsilon}}_{\mathrm{inc}}$ represents the complex polarization vector of $\mathbf{E}_{\mathrm{inc}}$, and the time dependence $\mathrm{exp}(-\mathrm{i}\omega t)$ has been dropped since it is the same for all fields. It is assumed that $\mathrm{Re}\{\cdot\}$ is taken to calculate the physical field, and therefore this notation is dropped too. We note that Eq.~(\ref{eqn: incidentE}) is valid for atomic media where the $z$-component of each mode $m_{i, z}$ is negligible (for the atomic systems considered in this paper, $m_{i, z} < 10^{-3}$ of the norm of the respective mode which is 1). Assuming they are not parallel, the modes therefore span the incident $\hat{\mathbf{x}}-\hat{\mathbf{y}}$ polarization plane. We also note that each term $c_{i}\hat{\mathbf{m}}_{i}$ still solves the wave equation since the coefficients $c_{i}$ are constant. When exiting the medium ($z = L$), the output field $\mathbf{E}_{\mathrm{out}}$ is then given by

\begin{equation}
\label{eqn: outputE}
\mathbf{E}_{\mathrm{out}}(z = L) = \mathcal{E}_{0}\big[ c_{1}t(n_{1})\hat{\mathbf{m}}_{1} + c_{2}t(n_{2})\hat{\mathbf{m}}_{2}\big]
\end{equation}

\noindent with $t(n_{i}) = \mathrm{exp}(\mathrm{i}n_{i}k_{0}L)$, where $L$ is the length of the medium. Therefore, each electric field mode only picks up a complex phase upon propagation. Since $n_{1} \neq n_{2}$, the medium is birefringent and dichroic, and therefore optically active. 

The only unknown variables in Eq.~(\ref{eqn: outputE}) are the complex coefficients $c_{i}$, and therefore we must calculate them. To do so, we use the initial matching condition of the electric field which is incident on the atomic vapor (at $L = 0$); this is Eq.~(\ref{eqn: incidentE}). Applying the inner product of each $\hat{\mathbf{m}}_{i}$ to both sides of Eq.~(\ref{eqn: incidentE}) gives two simultaneous equations which when recast as a matrix equation is

\begin{equation}
\begin{pmatrix} \langle\hat{\mathbf{m}}_{1}\vert\hat{\boldsymbol{\epsilon}}_{\mathrm{inc}}\rangle \\ \langle\hat{\mathbf{m}}_{2}\vert\hat{\boldsymbol{\epsilon}}_{\mathrm{inc}}\rangle 
\end{pmatrix} = \begin{pmatrix} \langle\hat{\mathbf{m}}_{1}\vert\hat{\mathbf{m}}_{1}\rangle & \langle\hat{\mathbf{m}}_{1}\vert\hat{\mathbf{m}}_{2}\rangle \\ \langle\hat{\mathbf{m}}_{2}\vert\hat{\mathbf{m}}_{1}\rangle & \langle\hat{\mathbf{m}}_{2}\vert\hat{\mathbf{m}}_{2}\rangle
\end{pmatrix}\begin{pmatrix} c_{1} \\ c_{2} 
\end{pmatrix} = \mathcal{S}\begin{pmatrix} c_{1} \\ c_{2} 
\end{pmatrix} \,,
\end{equation}

\noindent where the inner product of any two vectors $\mathbf{a}$ and $\mathbf{b}$ is defined as $\langle\mathbf{a}\vert\mathbf{b}\rangle = \mathbf{a}^{\dag}\cdot\mathbf{b}$, and the overlap matrix of the two modes is denoted $\mathcal{S}$. Since $\hat{\mathbf{m}}_{i}$ are normalized, the diagonal elements of $\mathcal{S}$ are $1$. In the special case of orthogonal modes, $\mathcal{S}$ is simply the identity matrix. The complex coefficient matrix is therefore 

\begin{equation} \label{eqn: complexCoefficients}
\begin{split}
\begin{pmatrix} c_{1} \\ c_{2} 
\end{pmatrix} & = 
\begin{pmatrix} 1 & \langle\hat{\mathbf{m}}_{1}\vert\hat{\mathbf{m}}_{2}\rangle \\ \langle\hat{\mathbf{m}}_{2}\vert\hat{\mathbf{m}}_{1}\rangle & 1
\end{pmatrix}^{-1}\begin{pmatrix} m_{1, x}^{*} & m_{1, y}^{*} & m_{1, z}^{*} \\ m_{2, x}^{*} & m_{2, y}^{*} & m_{2, z}^{*} 
\end{pmatrix}\begin{pmatrix} \epsilon_{\mathrm{inc}, x} \\ \epsilon_{\mathrm{inc}, y} \\ 0
\end{pmatrix} \,, \\ & = \mathcal{S}^{-1}M\hat{\boldsymbol{\epsilon}}_{\mathrm{inc}} \,,
\end{split}
\end{equation}

\noindent where $m_{i, j}$/$\epsilon_{\mathrm{inc}, j}$ are components of $\hat{\mathbf{m}}_{i}$/$\hat{\boldsymbol{\epsilon}}_{\mathrm{inc}}$ along each axis labeled $j$ in the lab frame, and the operation $M\hat{\boldsymbol{\epsilon}}_{\mathrm{inc}}$ is the projection of incident light polarization onto the mode matrix $M$. We also set $\epsilon_{\mathrm{inc}, z} = 0$ since the incident transverse light has no polarization along the propagation axis $\hat{\mathbf{z}}$. 
Note, Eq.~(\ref{eqn: complexCoefficients}) reinforces the terminology ``mode'' since each $\hat{\mathbf{m}}_{i}$ is invariant when it propagates through the medium (and therefore it maintains its shape). Specifically, one can calculate: 1) $c_{1} = 1$, $c_{2} = 0$ when $\hat{\boldsymbol{\epsilon}}_{\mathrm{inc}} = \hat{\mathbf{m}}_{1}$, and 2) $c_{1} = 0$, $c_{2} = 1$ when $\hat{\boldsymbol{\epsilon}}_{\mathrm{inc}} = \hat{\mathbf{m}}_{2}$. 

The complex coefficients $c_{i}$ are substituted into Eq.~(\ref{eqn: outputE}) to give the output electric field in the $\hat{\mathbf{x}}\hat{\mathbf{y}}\hat{\mathbf{z}}$ coordinate frame. We use $\mathbf{E}_{\mathrm{out}}$ to calculate the transmission of the incident electric field through the atomic vapor, following the Jones calculus method described in~\cite{keaveney2018elecsus, jones1941new}. It is worth noting that inside the medium, the modes---and therefore the electric field---are not transverse;
however, the displacement field $\mathbf{D}$ remains transverse. After exiting the medium, the electric field must be transverse in order to satisfy Maxwell's first equation in free space~\cite{griffiths2023introduction}. We assume that its $z$-component reflects off the interface boundary at $z = L$, and therefore we neglect it in all transmission calculations. Since we also assume, when writing Eq.~(\ref{eqn: incidentE}), that the $z$-component of each mode is negligible, setting ${E}_{\mathrm{out}, z} = 0$ has little consequence on the total output transmission.

\section{Methods}
\label{sec: method}

A non-zero mode overlap $\langle\hat{\mathbf{m}}_{i}\vert\hat{\mathbf{m}}_{j}\rangle$ ($i \neq j$) implies the electric field modes $\hat{\mathbf{m}}_{i}$ and $\hat{\mathbf{m}}_{j}$ are non-orthogonal. 
The degree of mode non-orthogonality, namely the value of $\langle\hat{\mathbf{m}}_{1}\vert\hat{\mathbf{m}}_{2}\rangle$, depends on $\theta_{B}$ and the other atomic parameters (in particular, $\chi$ is a function of $B$~\cite{zentile2015elecsus}). Consequently, the complex coefficients $c_{i}$, which are functions of $\hat{\mathbf{m}}_{i}$, are strongly dependent on these parameters. From Eq.~(\ref{eqn: complexCoefficients}), we can also tune the coefficients $c_{i}$ external to the atomic system by controlling the incident electric field polarization. In this paper, the tool we use to validate the new light propagation formalism---which accounts for mode non-orthogonality via $c_{i}$---is transmission spectroscopy, as shown in Fig.~\ref{fig: atomicSystem}(b). Light transmission $\tau$ quantifies the loss of light as it traverses the vapor cell, and is calculated using

\begin{equation} \label{eqn: tranmissionEqn}
\tau = \frac{\mathbf{E}_{\mathrm{out}}^{\dag}\mathbf{E}_{\mathrm{out}}}{{\mathbf{E}_{\mathrm{inc}}^{\dag}\mathbf{E}_{\mathrm{inc}}}} \,.
\end{equation}

\noindent In this study, $\mathbf{E}_{\mathrm{inc}}$ is sourced from a laser beam whose linear polarization lies in a plane inclined at an angle $\theta_{E}$ with respect to the $+\hat{\mathbf{x}}$ axis (see Fig.~\ref{fig: atomicSystem}). Varying $\theta_{E}$ allows us to control the incident light polarization $\hat{\boldsymbol{\epsilon}}_{\mathrm{inc}}$, and therefore tune the atom-light interaction via the coefficients $c_{i}$.

\begin{figure}
\centering
{\includegraphics[width=0.95\linewidth]{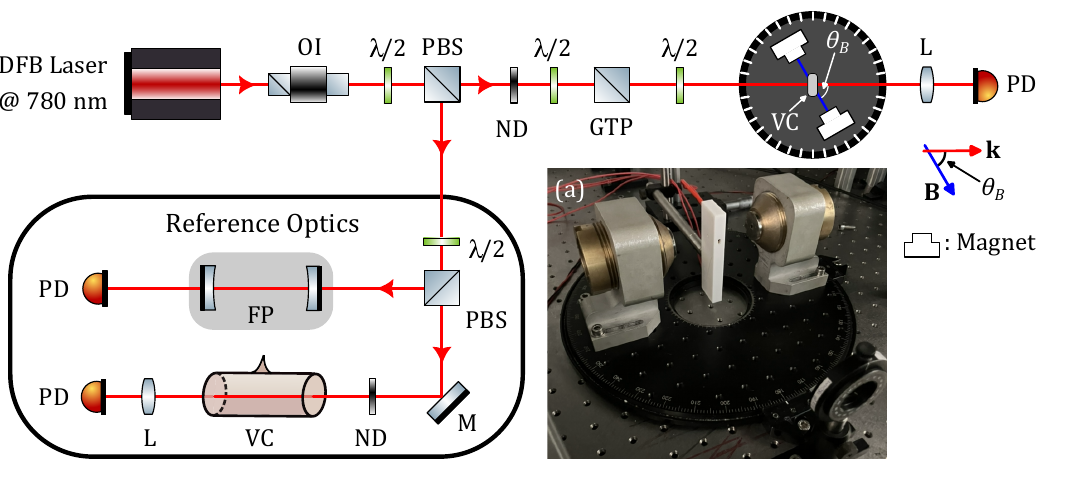}}
\caption{A schematic showing the experiment layout. A distributed feedback (DFB) diode laser which sweeps the Rb D$_{2}$ line is split between the main experiment and reference optics. The main experiment uses a Glan–Taylor polarizer (GTP) orientated to output linear light which is horizontally polarized. A half waveplate ($\lambda/2$) can rotate the light's polarization plane, before it traverses a Rb vapor cell (VC) of length $2~\mathrm{mm}$, which is situated in a resistive copper heater. The atoms are subject to an external magnetic field $\mathbf{B}$ produced by two NdFeB top-hat magnets, which are mounted in holders that can rotate by $\theta_{B}$ on a platform. Light transmission is measured using photodetectors (PD). Reference optics are used for data calibration~\cite{pizzey2022laser}, and consist of a Fabry-P\'erot etalon (FP) and a Rb VC atomic reference at room temperature. Optical power is controlled using neutral density filters (ND) and polarizing beamsplitter cubes (PBS). OI = optical isolator; M = mirror; L = lens. Picture (a): rotation platform, magnets, and VC holder/heater.}
\label{fig: experimentSetup}
\end{figure}

The experimental setup for transmission spectroscopy of a natural abundance Rb vapor cell is shown in Fig.~\ref{fig: experimentSetup}. Light is sourced from a distributed feedback (DFB) laser which sweeps the Rb D$_{2}$ transition (at wavelength $\lambda \approx 780~\mathrm{nm}$). The laser light is split between two arms: the reference arm, and the main experiment arm. The reference arm is used for data calibration, and consists of: 1) a Fabry-P\'erot etalon primarily to convert the laser scan to frequency, but also to correct for scan non-linearities; and 2) a natural abundance Rb vapor reference cell at room temperature, used to center data about a weighted line-center frequency (linear detuning $\Delta = 0~\mathrm{GHz}$)~\cite{pizzey2022laser}. Along the main experimental arm, the initial polarization is defined using a Glan-Taylor polarizer (GTP); after the GTP, light has horizontal-linear polarization ($\theta_{E} = 0^{\circ}$). Prior to the same GTP, a half waveplate (HWP) and neutral density filters are used to control optical power to the transmission spectroscopy experiment. After the GTP, a further HWP controls the plane of linear polarization, before the light traverses a Rb vapor cell of length $L = 2~\mathrm{mm}$. The cell is situated in a resistive copper heater, enclosed in a Teflon cover for thermal shielding. We fix the vapor cell within its heater at the center of a Thorlabs rotating breadboard (RBB300A/M), that is used to rotate two NdFeB top-hat permanent magnets. 
The wavevector $\mathbf{k}$ of the incident light is thus fixed in the $z$-direction while the external magnetic field is rotated by $\theta_{B}$ with respect to $\mathbf{k}$. This matches the assumptions set when constructing our model. By adjusting the magnet separation, we can span fields with a magnitude as little as hundreds of Gauss, up to $\approx 6~\mathrm{kG}$. For larger magnetic fields, a $2~\mathrm{mm}$ cell ensures field homogeneity along both $\hat{\mathbf{x}}$ and $\hat{\mathbf{z}}$. At the center of the platform, the beam waist has size $(426 \pm 1)~\mathrm{\upmu m} \times (634 \pm 2)~\mathrm{\upmu m}$, which was measured using a Thorlabs CMOS camera~\cite{keaveney2018automated} and custom image analysis code. The optical power is set to order $100~\mathrm{nW}$, which ensures the experiment is in the weak-probe regime~\cite{sherlock2009weak, siddons2008absolute, haupl2025modelling}. After light traverses the vapor, we measure its transmission as a voltage using a photodetector, which is recorded on an oscilloscope. Note, assuming Gaussian laser beam propagation~\cite{adams2018optics}, the spatial angular spread of the laser in our experiment is $0.03^{\circ}$; the plane-wave approximation used in the theory section is therefore valid.

\section{Results}
\label{sec: results}

Having established a technique, we now explore two distinct regimes that explicitly show non-orthogonal modes are predicted by the wave equation. In both regimes, experimental data is taken to validate that mode non-orthogonality is correctly accounted for by the new light propagation formalism. The first regime we consider---denoted \emph{Regime I}---is where the magnitude of the magnetic field is relatively small, and the magnetic field angle is near the Voigt geometry ($\theta_{B} = \pi/2$, where modes are orthogonal). This regime has generated a recent burgeoning of interest due to the significant improvement to atomic filter performance~\cite{rotondaro2015generalized, keaveney2018elecsus, higgins2020atomic, alqarni2024device}. We use ``small'' to describe the regime $B << A_{\mathrm{hf}}/\mu_{\mathrm{B}} = B_{\mathrm{HPB}}$, where $A_{\mathrm{hf}}$ is the ground state magnetic dipole constant, $\mu_{\mathrm{B}}$ is the Bohr magneton, and $B_{\mathrm{HPB}}$ approximates a magnetic field where the Zeeman effect is equivalent to the hyperfine interaction~\cite{pizzey2022laser}. For $^{87}$Rb, $B_{\mathrm{HPB}} \approx 2.4~\mathrm{kG}$; at this field, the atoms \emph{begin} to enter the hyperfine Paschen-Back (HPB) regime~\cite{sargsyan2012hyperfine, zentile2014hyperfine}. In the HPB regime where $B >> B_{\mathrm{HPB}}$---denoted \emph{Regime III}---ground atomic eigenstates strongly decouple into the $\vert m_{I}, m_{J}\rangle$ basis ($m_{I}$, $m_{J}$ are projection quantum numbers), and atomic transitions separate into resolvable groups based on electric-dipole transition types~\cite{ponciano2020absorption, reed2018low, briscoe2023voigt, briscoe2024indirect}. The HPB regime therefore offers a much simpler atomic system for thermal vapor studies~\cite{ciampini2017optical, staerkind2023precision, ponciano2020absorption, mottola2023electromagnetically}, but is not as interesting for testing new theory. For the second regime, we choose $B \approx B_{\mathrm{HPB}}$ and $\theta_{B} = 130^{\circ}$ (near halfway between the Faraday and Voigt geometries)---denoted \emph{Regime II}. In this intermediate regime, $B$ is not so large to completely enter the HPB regime. For $B \approx B_{\mathrm{HPB}}$, we rely on a matrix representation of the atomic Hamiltonian to calculate transition frequencies and line-strengths~\cite{zentile2015elecsus}. Furthermore, $\theta_{B} = 130^{\circ}$ means there is close to equal projection of the polarization plane onto the axes parallel and perpendicular to the atomic quantization axis. In theory, all types of electric-dipole transition can therefore be driven ($\pi$, $\sigma^{+}$ and $\sigma^{-}$), depending on the atom-light coupling. The more complex resonance structure at $B \approx B_{\mathrm{HPB}}$, in tandem with the non-trivial atom-light interaction for $\theta_{B} = 130^{\circ}$, makes it an excellent regime for testing any new theory. A range of transmission heatmaps (theory) are plotted in the supplemental document which highlight disagreement between the old and new light propagation formalisms as a continuous function of $\theta_{B}$ and $\theta_{E}$. These extend the atom-light parameter space explored in this results section.

\subsection{Regime I}
\label{subSec: regime1Modeling}

\subsubsection{Modeling}
\label{subSubSec: regime1Modeling}

\begin{figure}
\centering
{\includegraphics[width=0.975\linewidth]{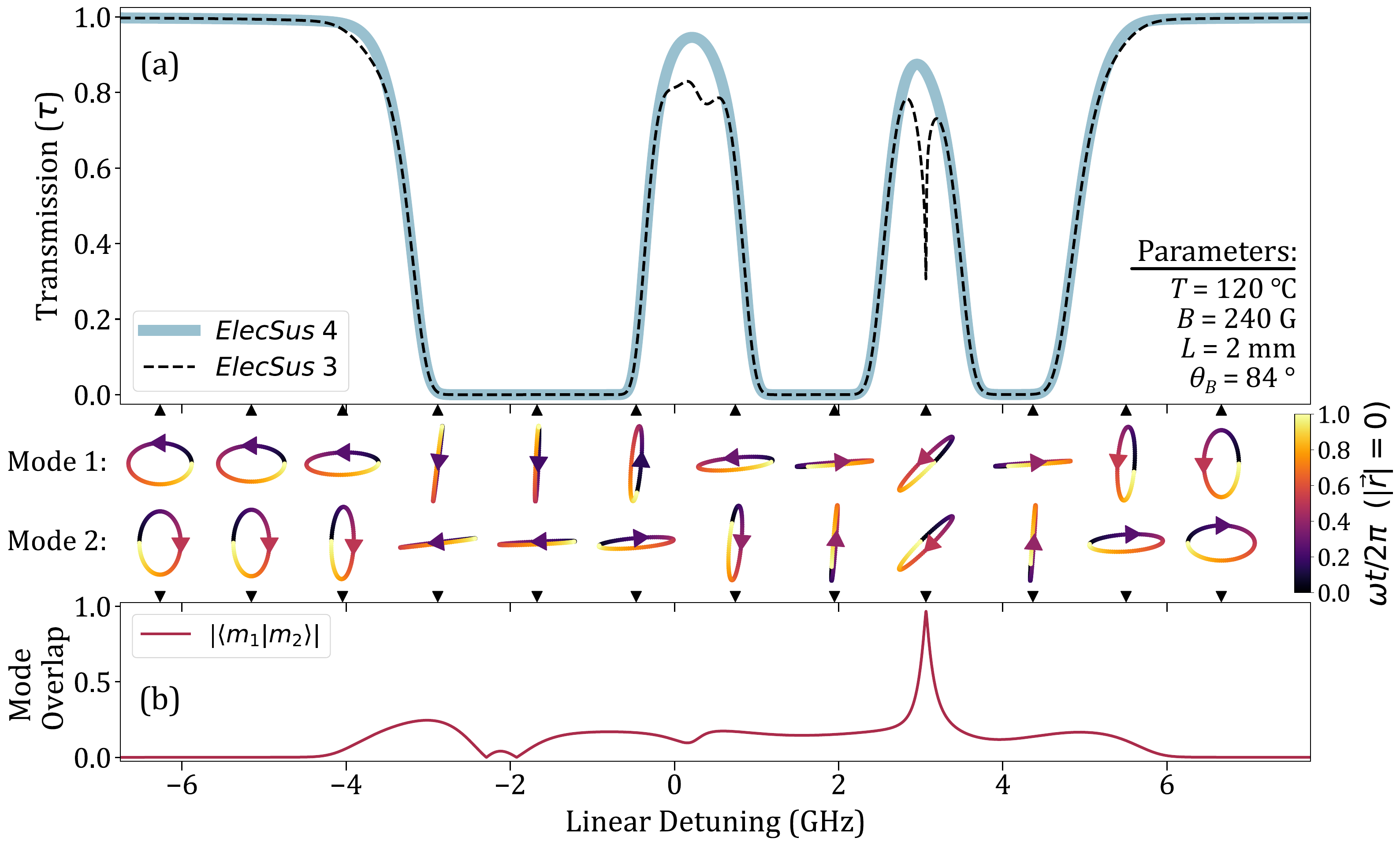}}
\caption{Theoretical transmission spectra ($\tau$) for a Rb atomic vapor cell of natural abundance on the D$_{2}$ line, which is subject to horizontal-linear light. The atomic system parameters are cell length $L$, atom temperature $T$, magnetic field magnitude $B = \lvert\mathbf{B}\rvert$, and the angle $\theta_{B}$ between $\mathbf{B}$ and the light wavevector $\mathbf{k}$. (a) prediction of \emph{ElecSus}~$4$ (blue), which accounts for an atomic medium that exhibits non-orthogonal electric field modes, plotted against the old model \emph{ElecSus}~$3$~\cite{keaveney2018elecsus} (black, dashed), which does not. (b) magnitude of the electric field mode overlap $\vert\langle \hat{\mathbf{m}}_{1}\vert \hat{\mathbf{m}}_{2}\rangle\vert$ (red), highlighting both the frequency dependence and non-orthogonality of the modes $\hat{\mathbf{m}}_{i}$. Between (a) and (b), each mode is plotted as a real-space polarization ellipse whose phase evolves in time $t$. These ellipses are centered at unique frequencies, indicated by black triangles which point to the linear detuning axis. Both the color and arrow direction of each ellipse are used to indicate mode handedness.}
\label{fig: nonOrthogonalModesSmallB}
\end{figure}

The first region of atom-light parameter space we explore is Regime I. In Fig.~\ref{fig: nonOrthogonalModesSmallB}(a), we plot light transmission through a natural abundance Rb vapor ($\approx 27\%$ $^{87}$Rb) with the following atom-light system parameters: D-line = D$_{2}$, vapor cell length $L = 2~\mathrm{mm}$, atom temperature $T = 120~^{\circ}\mathrm{C}$, $B = 240~\mathrm{G}$, $\theta_{B} = 84^{\circ}$, and $\theta_{E} = 0^{\circ}$ (i.~e.~$\hat{\boldsymbol{\epsilon}}_{\mathrm{inc}} = [1, 0, 0]^\mathrm{T}$). The predicted transmission is plotted using both \emph{ElecSus}~$4$ (blue) and \emph{ElecSus}~$3$ (black, dashed). Significant disagreement between models is seen at linear detunings $\Delta \approx 0~\mathrm{GHz}$ and $\Delta \approx 3~\mathrm{GHz}$; elsewhere, the models mostly agree well. The \emph{ElecSus}~$3$ spectrum feature at $\Delta \approx 3~\mathrm{GHz}$ is most notable---and unphysical---for a light transmission spectrum. This is because $\tau$ depends on light attenuation, a function of Doppler-broadened symmetric Voigt profiles centered on atomic resonances, each with a Gaussian dominated full width at half maximum at $100~^{\circ}\mathrm{C}$ of $577~\mathrm{MHz}$~\cite{pizzey2022laser} ($^{85}$Rb D$_{2}$). It is clear from Fig.~\ref{fig: nonOrthogonalModesSmallB}(a) that this feature does not resemble those properties.

To understand any disagreement between the models, we need a mechanism to visualize the degree of mode non-orthogonality as a function of frequency, as this is what differentiates the old and new light propagation formalisms; in Fig.~\ref{fig: nonOrthogonalModesSmallB}, we show two ways. In panel (b), we gain insight into the effect of the overlap matrix components $\langle\hat{\mathbf{m}}_{1}\vert\hat{\mathbf{m}}_{2}\rangle = \langle\hat{\mathbf{m}}_{2}\vert\hat{\mathbf{m}}_{1}\rangle^{*}$ by plotting $\vert\langle\hat{\mathbf{m}}_{1}\vert \hat{\mathbf{m}}_{2}\rangle\vert$, which we call the degree of mode overlap. It is clear from panel (b) that in the region of strongest $\vert\langle \hat{\mathbf{m}}_{1}\vert \hat{\mathbf{m}}_{2}\rangle\vert$, we see the most significant disagreement between the old and new light propagation formalisms. In this example, the disagreement is as large as $55\%$ of the total transmission. We note that the complex coefficients $c_{i}$ are calculated using both $\langle \hat{\mathbf{m}}_{1}\vert \hat{\mathbf{m}}_{2}\rangle$ and $\hat{\boldsymbol{\epsilon}}_{\mathrm{inc}}$, not $\vert\langle \hat{\mathbf{m}}_{1}\vert \hat{\mathbf{m}}_{2}\rangle\vert$, meaning there is not a one-to-one correspondence between $\tau$ and $\vert\langle \hat{\mathbf{m}}_{1}\vert \hat{\mathbf{m}}_{2}\rangle\vert$; we use the degree of mode overlap only to gain insight into the physical magnitude of the overlap $\langle \hat{\mathbf{m}}_{1}\vert \hat{\mathbf{m}}_{2}\rangle$, which is a complex quantity. The shape and phase of the modes can instead be represented as real-space polarization ellipses, as shown between panels (a) and (b) of Fig.~\ref{fig: nonOrthogonalModesSmallB} for various frequencies $\Delta_{k}$ (indicated by black triangles which point to the linear detuning axis). Each ellipse is plotted as $\mathrm{Re}[m_{i, y}(\Delta_{k})\mathrm{exp}(-\mathrm{i}\omega t)]$ vs $\mathrm{Re}[m_{i, x}(\Delta_{k})\mathrm{exp}(-\mathrm{i}\omega t)]$, where the mode $\hat{\mathbf{m}}_{i}(\Delta_{k}) \approx [m_{i, x}(\Delta_{k}), m_{i, y}(\Delta_{k}), 0]^{\mathrm{T}}\mathrm{exp}(-\mathrm{i}\omega t)$ evolves in time $t$ between $t = 0$ and $t = 2\pi/\omega$, and we have assumed a negligible $z$-component for each mode. Mode handedness is represented by both a colormap and an arrow. It is very clear from these ellipse plots, and the degree of mode overlap, that the electric field mode solutions to the wave equation are both frequency dependent and non-orthogonal. We therefore must use the new formalism for light propagation to account for non-orthogonal modes.

\subsubsection{Experiment}
\label{subSubSec: regime1Experiment}

\begin{figure}
\centering
{\includegraphics[width=\linewidth]{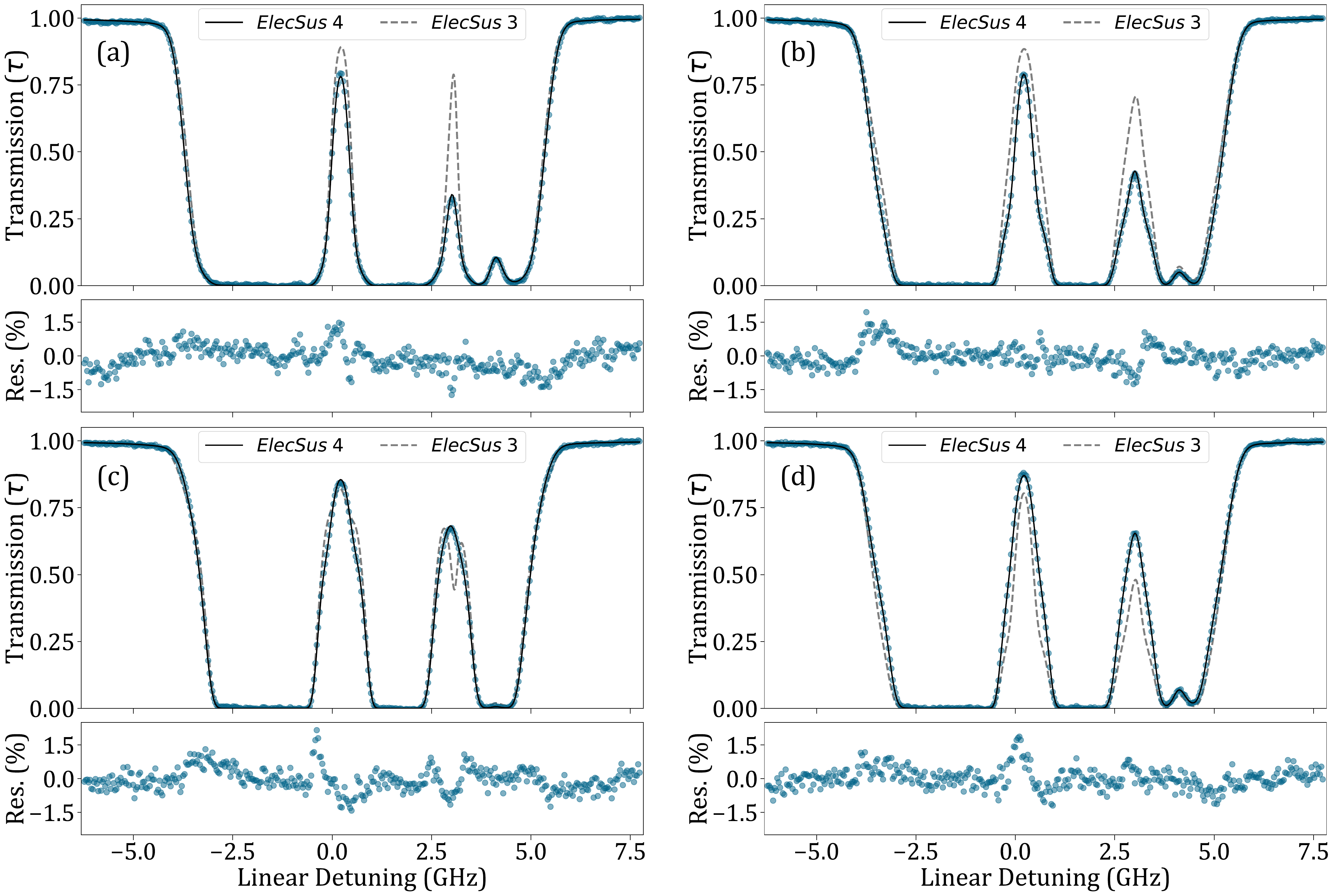}}
\centering
\caption{Experimental transmission spectroscopy of a natural abundance Rb vapor cell of length $L = 2~\mathrm{mm}$. The atoms are subject to an external magnetic field oriented at $\theta_{B} = 80^{\circ}$. Panels (a)-(d) show the data (blue) of spectra for 4 different incident linear polarization angles: (a) $\theta_{E} = (90.9 \pm 0.5)^{\circ}$, (b) $\theta_{E} = (49.1 \pm 0.1)^{\circ}$, (c) $\theta_{E} = (19.8 \pm 0.2)^{\circ}$, and (d) $\theta_{E} = (-48.1 \pm 0.1)^{\circ}$. These angles were found by simultaneously fitting the data to \emph{ElecSus}~$4$. The following global parameters of all spectra were also extracted by the fit: $T = (119.69 \pm 0.05)~^{\circ}\mathrm{C}$, $B = (239.0 \pm 0.2)~\mathrm{G}$ and $\Gamma_{\mathrm{B}} = (7.9 \pm 0.2)~\mathrm{MHz}$. The fit parameters extracted using \emph{ElecSus}~$4$ are used to plot theory curves with both \emph{ElecSus}~$4$ (black) and \emph{ElecSus}~$3$ (gray, dashed). Residuals quantify deviation of data from the \emph{ElecSus}~$4$ fit, and show excellent agreement.}
\label{fig: lowBfieldExperiment}
\end{figure}

The first experiment we performed was in Regime I, analogous to Fig.~\ref{fig: nonOrthogonalModesSmallB}. We fix the magnetic field angle $\theta_{B}$, while the incident plane of linear polarization was rotated via a HWP. This allows us to investigate the effect of changing the atom-light coupling via tuning the complex coefficients $c_{i}$ for a fixed atomic system. The magnetic field was set such that $\theta_{B} = 80^{\circ}$ in order to probe the atomic system for unphysical features predicted by the old light propagation formalism (analogous to those in Fig.~\ref{fig: nonOrthogonalModesSmallB}). The HWP was calibrated external to the experiment in order to set a linear polarization angle $\theta_{E}$. In Fig.~\ref{fig: lowBfieldExperiment}, we plot the data (blue) of transmission spectra for four different incident linear polarization angles; at each $\theta_{E}$, a transmission spectrum was recorded 4 times (repeats) to account for random errors~\cite{hughes2010measurements}. The data of each spectrum were processed using a similar method to~\cite{pizzey2022laser}, and fit to \emph{ElecSus}~$4$. Since the only variable changed during the experiment was $\theta_{E}$, each spectrum should have the same atom parameters $T$, $B$, $\theta_{B}$, and additional Lorentzian broadening $\Gamma_{\mathrm{B}}$ (e.~g.~from the presence of a buffer gas in the vapor cell~\cite{rotondaro1997collisional}). The atom temperature may vary slightly over the experiment because the $2~\mathrm{mm}$ vapor cell was not actively temperature stabilized. When fitting the data, we employed a simultaneous fitting algorithm analogous to the technique used in~\cite{briscoe2024indirect}, adding constraints so that the parameters $T$, $B$, $\theta_{B}$ and $\Gamma_{\mathrm{B}}$ were required to fit to a global value for every transmission spectrum. For global parameters, the fit returned: $T = (119.69 \pm 0.05)~^{\circ}\mathrm{C}$, $B = (239.0 \pm 0.2)~\mathrm{G}$, $\Gamma_{\mathrm{B}} = (7.9 \pm 0.2)~\mathrm{MHz}$, and $\theta_{B} = (80 \pm 0.5)^{\circ}$. These were found by taking the mean and standard error across the fit parameters extracted separately from each repeat~\cite{hughes2010measurements}. We note that $\theta_{B}$ was set to its measured value in the fit, and $\alpha_{\theta_{B}} = 0.5^{\circ}$ is an approximate measurement error. This reduces the complexity of the simultaneous fit, and also ensures that the most important atomic parameter which dictates mode non-orthogonality matches the lab measurement. For the spectra shown in Fig.~\ref{fig: lowBfieldExperiment}, the polarization angles were: (a) $\theta_{E} = (90.9 \pm 0.5)^{\circ}$, (b) $\theta_{E} = (49.1 \pm 0.1)^{\circ}$, (c) $\theta_{E} = (19.8 \pm 0.2)^{\circ}$, and (d) $\theta_{E} = (-48.1 \pm 0.1)^{\circ}$. The majority of the global fit parameters were within $1-2\%$ percentage error from the expected value, based on measurements and calibrations taken in the lab. The main deviation in our results was $T$, which was approximately $3~^{\circ}\mathrm{C}$ lower than measured ($2.6\%$ percentage error). We attribute this to measuring the copper block temperature, which is not completely in contact with the vapor cell stem (whose temperature sets the atom temperature). Similar temperature deviations have been reported in literature~\cite{hanley2015absolute}, in some cases as large as $10~^{\circ}\mathrm{C}$~\cite{uhland2023build}. For $\theta_{E}$, the angular deviation of the fit from calibration was: (a) $1.6^{\circ}$, (b) $-0.2^{\circ}$, (c) $0.4^{\circ}$, and (d) $2.6^{\circ}$, which again shows good agreement. In Fig.~\ref{fig: lowBfieldExperiment}, we use the simultaneous fit parameters extracted using \emph{ElecSus}~$4$ to plot the predicted \emph{ElecSus}~$4$ (black) and \emph{ElecSus}~$3$ spectra (gray, dashed). We do not fit to \emph{ElecSus}~$3$, since we found this returns either: 1) bad lineshapes fit to parameters which do not match those measured in the lab, or 2) even worse lineshapes if parameters such as $\theta_{B}$ are constrained in the fit to match the measured lab value. For completeness, we show an example of \emph{ElecSus}~$3$ fitting in the supplemental document (using the $\theta_{E} = 90.9^{\circ}$ data). \emph{ElecSus}~$3$ theory plots are shown in Fig.~\ref{fig: lowBfieldExperiment} to highlight the disagreement between models at the extracted \emph{ElecSus}~$4$ fit parameters. The new light propagation formalism clearly fits with excellent agreement to data taken in the lab, fitting to \emph{ElecSus}~$4$ with a reduced mean square error (RMSE) of $0.7\%$ for each spectra in Fig.~\ref{fig: lowBfieldExperiment} (taken over the range $\pm 7~\mathrm{GHz}$). We conclude that the data taken in Regime I fits with excellent agreement to the new light propagation formalism that accounts for non-orthogonal electric field modes.

\subsection{Regime II}
\label{subSec: regime2}

\subsubsection{Modeling}
\label{subSubSec: regime2Modeling}

\begin{figure}
\centering
{\includegraphics[width=0.975\linewidth]{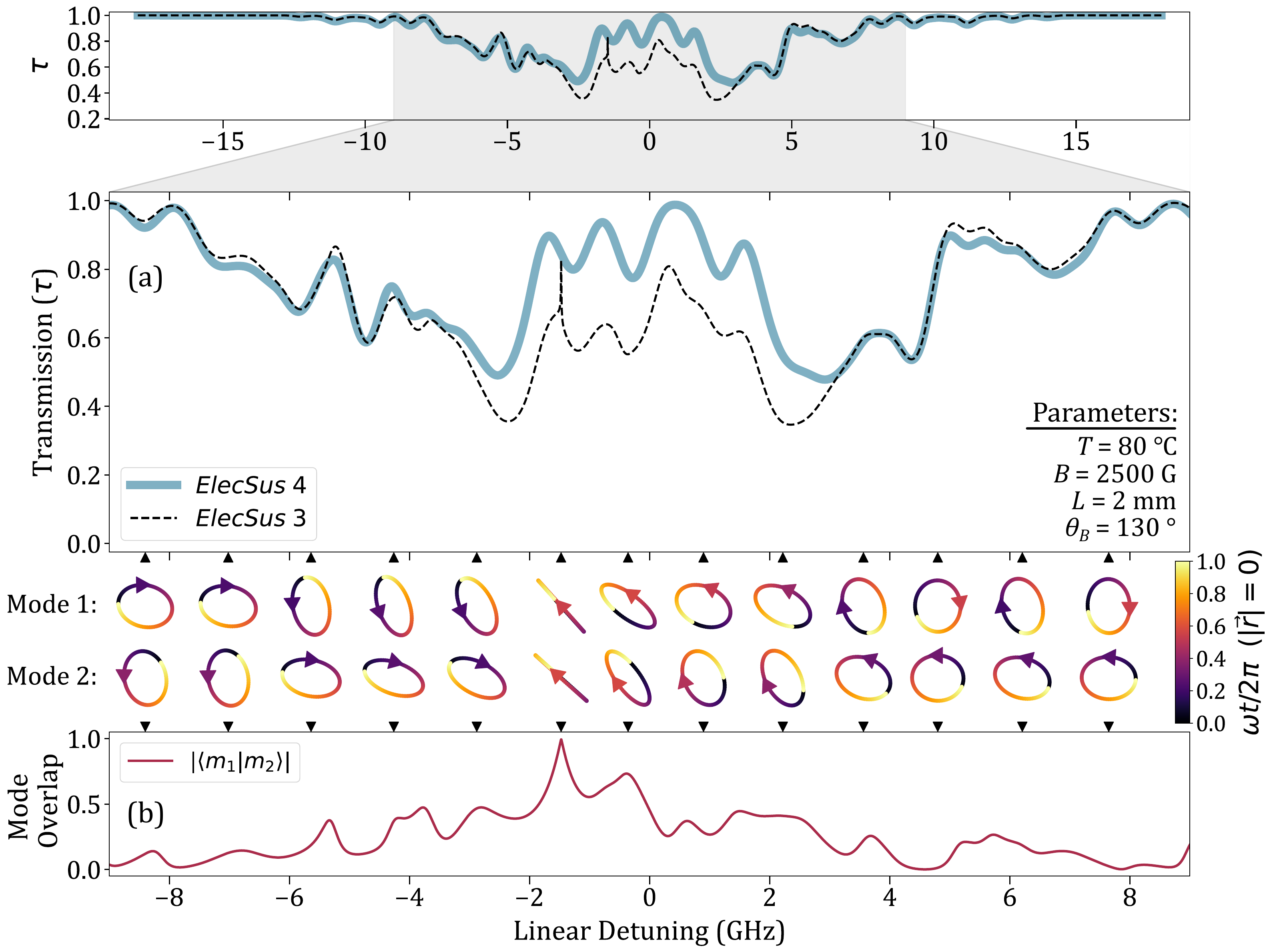}}
\caption{Theoretical transmission spectra ($\tau$) for a Rb atomic vapor cell of natural abundance on the D$_{2}$ line, which is subject to vertical-linear light. The figure follows the same structure as Fig.~\ref{fig: nonOrthogonalModesSmallB} for the atomic parameters shown within the figure. A larger frequency range is shown in the additional panel at the top.}
\label{fig: nonOrthogonalModesLargeB}
\end{figure}

The second region we explore is Regime II. Fig.~\ref{fig: nonOrthogonalModesLargeB} shows the theoretical transmission spectrum of a naturally abundant Rb vapor with the following parameters: D-line = D$_{2}$, $L = 2~\mathrm{mm}$, $T = 80~^{\circ}\mathrm{C}$, $B = 2.5~\mathrm{kG}$, $\theta_{B} = 130^{\circ}$ and $\theta_{E} = 90^{\circ}$ (i.~e.~$\hat{\boldsymbol{\epsilon}}_{\mathrm{inc}} = [0, 1, 0]^\mathrm{T}$). The figure follows the same structure as Fig.~\ref{fig: nonOrthogonalModesSmallB}, with an additional panel to show the full frequency range of the spectrum. The main plot focuses on the region of strongest mode overlap. Regime I showed only narrow frequency ranges where \emph{ElecSus}~$3$ and \emph{ElecSus}~$4$ outputs disagreed; in Regime II, there is strong disagreement over a much larger frequency range. The disagreement is again accompanied by a corresponding frequency range where the degree of mode overlap is large, and mode ellipse plots which are visibly non-orthogonal. As was the case in Regime I, we note an unphysical feature at $\Delta \approx -1.5~\mathrm{GHz}$ in Fig.~\ref{fig: nonOrthogonalModesLargeB}(a), which should not be predicted in a Doppler-broadened light transmission spectrum. It is clear that at frequencies where the electric field modes exhibit non-orthogonality, the new light propagation formalism significantly disagrees with the old formalism.

\subsubsection{Experiment}
\label{subSubSec: regime2Experiment}

\begin{figure}
\centering
{\includegraphics[width=0.95\linewidth]{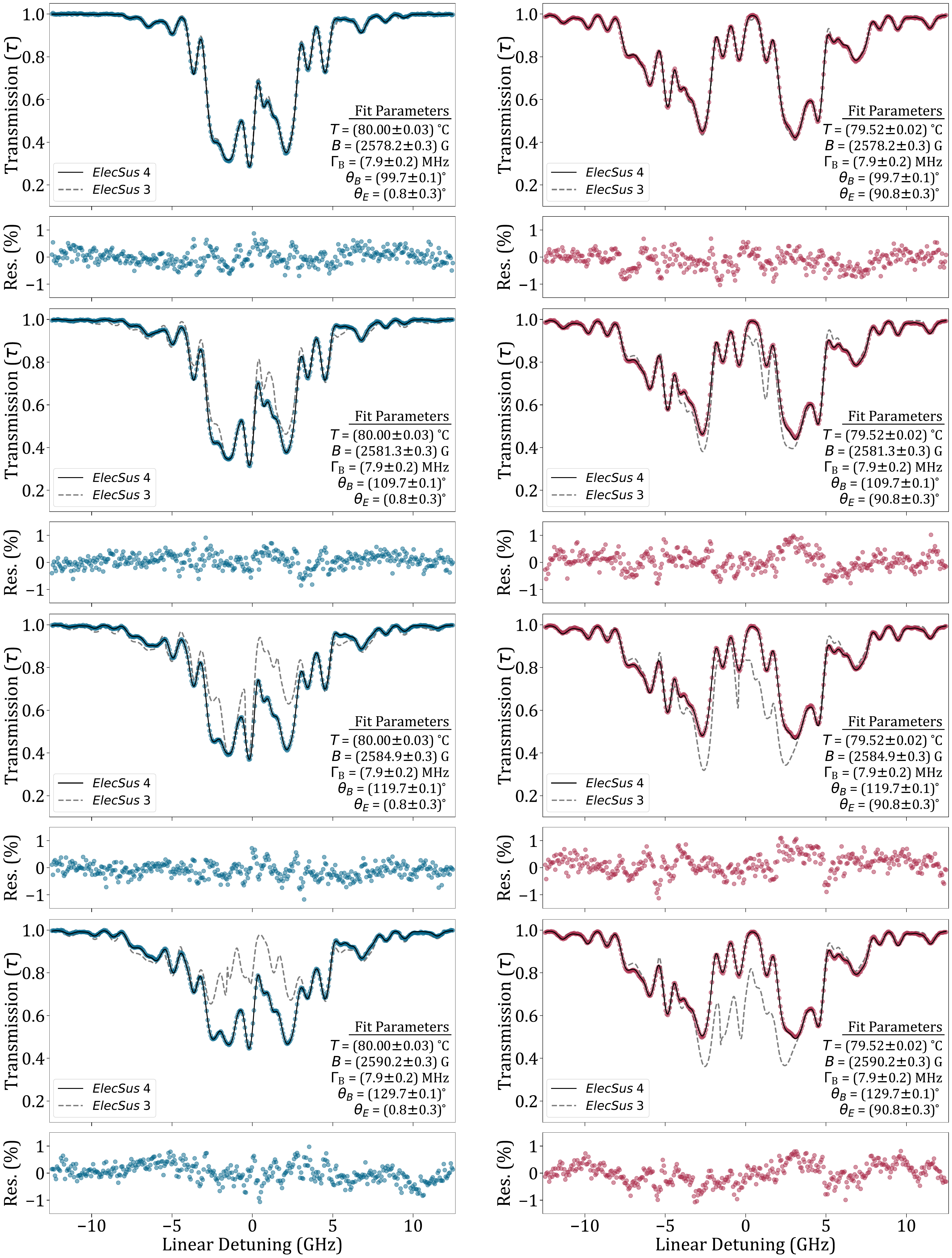}}
\caption{Experimental transmission spectroscopy of a natural abundance Rb vapor cell of length $L = 2~\mathrm{mm}$. The atoms are subject to a large external magnetic field, oriented at $\theta_{B}$ with respect to the light wavevector $\mathbf{k} = k\hat{\mathbf{z}}$, and linearly polarized light, oriented at $\theta_{E}$ with respect to $\hat{\mathbf{x}}$. Data is plotted for spectra where $\theta_{E} = 0.8^{\circ}$ (blue) and $\theta_{E} = 90.8^{\circ}$ (red). The following parameters were extracted by fitting the data to \emph{ElecSus}~$4$: $T$, $B$, and $\theta_{B}$. The parameters $\Gamma_{\mathrm{B}}$ and $\theta_{E}$ were calibrated using the Regime I experiment (see Fig.~\ref{fig: lowBfieldExperiment}). Panels are labeled with the parameters of the spectrum within: each column has a different $\theta_{E}$, and each row has a different $\theta_{B}$. Labeled parameters are used to plot theory curves with both \emph{ElecSus}~$4$ (black) and \emph{ElecSus}~$3$ (gray, dashed). Residuals quantify deviation of data from the \emph{ElecSus}~$4$ fit, and show excellent agreement.}
\label{fig: highBfieldExperiment}
\end{figure}

For the Regime II experiment, we fix the incident light polarization while varying the magnetic field angle. By varying $\theta_{B}$, we change the atomic system, and therefore the wave equation will have different unique solutions. The shape of the modes will therefore change, with a concomitant tuning of the atom-light coupling via the complex coefficients $c_{i}$. We chose two fixed polarizations; these were horizontal-linear light---$\theta_{E} = (0.8\pm0.3)^{\circ}$---and vertical-linear light---$\theta_{E} = (90.8\pm0.3)^{\circ}$. The linear polarization angles $\theta_{E}$ were extracted using the HWP calibration from the Regime I experiment. For each fixed polarization, we rotate the magnetic field in steps of $\Delta\theta_{B} = 10^{\circ}$ between $\theta_{B} = 100^{\circ}$ and $\theta_{B} = 130^{\circ}$. In the Voigt geometry ($\theta_{B} = 90^{\circ}$), the atom-light system exhibits orthogonal electric field modes; we therefore expect zero disagreement between \emph{ElecSus}~$3$ and \emph{ElecSus}~$4$ outputs. As $\theta_{B}$ is rotated away from $90^{\circ}$, the modes exhibit a stronger degree of frequency-dependent non-orthogonality; the new light propagation formalism should therefore disagree more with the old formalism. This is exactly what we saw in our experiment, shown in Fig.~\ref{fig: highBfieldExperiment}. The figure is structured as follows: (i) two columns, which differ by $\theta_{E}$; and (ii) four rows, where $\theta_{B}$ increases in steps of $10^{\circ}$. A simultaneous fitting routine was again employed to fit the data, constraining global parameters in the fit. The fit parameters are added to each panel in Fig.~\ref{fig: highBfieldExperiment}; fit constraints can be identified by comparing the parameters in each row and column (which have the same value). For clarity, datasets where horizontal-linear light was incident are plotted in blue, whereas datasets where vertical-linear light was incident are plotted in red. Since the rotation platform is large, we added a fit constraint that while $\theta_{B}$ may float, each $\theta_{B}$ must increase by steps of $10^{\circ}$ from the last. For each $\theta_{B}$, we constrained $B$ for both incident polarizations; the field deviation $\Delta B \approx 10~\mathrm{G}$ over a $\Delta\theta_{B} = 30^{\circ}$ angular range is $<0.5\%$ of $B$, which suggests excellent alignment of the cell between the magnets. The values of $\Gamma_{\mathrm{B}}$ and calibrated $\theta_{E}$ (based on the HWP angle used) were fed through from the Regime I experiment, and fixed in the fit to reduce its complexity. All fit parameters were within $2\%$ percentage error of the expected value, based on measurements and calibrations taken in the lab. In Fig.~\ref{fig: highBfieldExperiment}, we use the simultaneous fit parameters to plot the predicted \emph{ElecSus}~$4$ (black) and \emph{ElecSus}~$3$ spectra (gray, dashed). As was the case in Regime I, \emph{ElecSus}~$3$ theory plots are shown to highlight the degree of disagreement between models at the extracted fit parameters. Considering spectra with $\theta_{E} = (0.8\pm0.3)^{\circ}$ only, the RMSE of each spectra taken over the range $\pm 14~\mathrm{GHz}$ was $0.5\%$ using \emph{ElecSus}~$4$.  Note, although the $\theta_{E} = 90.8^{\circ}$ datasets in Fig.~\ref{fig: highBfieldExperiment} seem insensitive to $\theta_{B}$, there is nonetheless a variation and this variation is sufficient for \emph{ElecSus}~$4$ to find $\theta_{B}$ accurately. We conclude that the data in Regime II fit with excellent agreement to a model that accounts for non-orthogonal modes. Both experiments cover a large atomic parameter space, and demonstrate excellent agreement between data and theory. We therefore conclude that the correct light propagation formalism must properly account for an atomic system which exhibits non-orthogonal electric field modes.

\section{Conclusions}
In this paper, we showed an atomic vapor exhibits non-orthogonal electric field modes when subject to an external magnetic field that is oblique to the light propagation direction. Mode non-orthogonality is not properly accounted for by atomic vapor models in the literature, and consequently, large unphysical features are predicted by existing modeling programs \cite{keaveney2018elecsus} in light transmission spectra. We therefore derived a new light propagation formalism to account for atomic systems which exhibit non-zero mode overlap. The unphysical features are not only removed by the new propagation formalism, but the new theory also predicts light transmission through an atomic vapor with excellent agreement to experiment. For a natural abundance Rb vapor cell on the D$_{2}$ line, we showed this was not only true for both small and large magnetic field magnitudes, but also across a wide range of magnetic field angles relative to the light wavevector. We showed that the new light propagation formalism can correct unphysical transmission features with greater than $50\%$ disagreement to the old formalism over narrow frequency ranges of $\approx 1-2~\mathrm{GHz}$. We also showed a regime where unphysical features which span much larger frequency ranges of $\approx 10~\mathrm{GHz}$ were corrected too. In conclusion, 
this work demonstrates conclusively that the correct light propagation formalism for an atomic vapor subject to a magnetic field of arbitrary direction must account for the possibility that the electric field modes are non-orthogonal. The frequency dependence of the modes, which was verified indirectly in this article, will be explored in future work as a novel mechanism to suppress the bandwidth of single-cell atomic filters, as shown experimentally in~\cite{keaveney2018optimized}. We also aim to explore possible research avenues which exploit exceptional points of degeneracy in a simple thermal vapor experiment where non-orthogonal modes completely coalesce~\cite{logueThesis}. Given the analytic nature of our model, any predictions can be optimized and experimentally verified.

\begin{backmatter}
\bmsection{Funding}
Engineering and Physical Sciences Research 
 Council (EP/T518001/1 \& EP/R002061/1)

\bmsection{Acknowledgments}
The authors thank Lucy A Downes, Tobias Franzen and Liam P Gallagher for valuable discussions.

\bmsection{Disclosures}
The authors declare no conflicts of interest.

\bmsection{Data availability} Data underlying the results presented in this paper are available in~\cite{data}.

\bmsection{Supplemental document}
See supplemental document for supporting information referenced in the paper.

\end{backmatter}

\bibliography{bibliography}

\begin{thebibliography}{10}
\newcommand{\enquote}[1]{``#1''}

\bibitem{fan2015atom}
H.~Fan, S.~Kumar, J.~Sedlacek, \emph{et~al.}, \enquote{{Atom based RF electric field sensing},} {\protect\JournalTitle{Journal of Physics B: Atomic, Molecular and Optical Physics}} \textbf{48}, 202001 (2015).

\bibitem{allinson2023simultaneous}
G.~Allinson, M.~J. Jamieson, A.~R. Mackellar, \emph{et~al.}, \enquote{{Simultaneous multi-band radio-frequency detection using high-orbital-angular-momentum states in a Rydberg-atom receiver},} {\protect\JournalTitle{Physical Review Research}} \textbf{6}, 023317 (2024).

\bibitem{jau2020vapor}
Y.-Y. Jau and T.~Carter, \enquote{{Vapor-cell-based atomic electrometry for detection frequencies below 1 kHz},} {\protect\JournalTitle{Physical Review Applied}} \textbf{13}, 054034 (2020).

\bibitem{downes2020full}
L.~A. Downes, A.~R. MacKellar, D.~J. Whiting, \emph{et~al.}, \enquote{{Full-field terahertz imaging at kilohertz frame rates using atomic vapor},} {\protect\JournalTitle{Physical Review X}} \textbf{10}, 011027 (2020).

\bibitem{knappe2006microfabricated}
S.~Knappe, P.~Schwindt, V.~Gerginov, \emph{et~al.}, \enquote{{Microfabricated atomic clocks and magnetometers},} {\protect\JournalTitle{Journal of Optics A: Pure and Applied Optics}} \textbf{8}, S318 (2006).

\bibitem{gharavipour2016high}
M.~Gharavipour, C.~Affolderbach, S.~Kang, \emph{et~al.}, \enquote{{High performance vapour-cell frequency standards},} in \emph{Journal of Physics: Conference Series,}  vol. 723 (IOP Publishing, 2016), p. 012006.

\bibitem{lvovsky2009optical}
A.~I. Lvovsky, B.~C. Sanders, and W.~Tittel, \enquote{{Optical quantum memory},} {\protect\JournalTitle{Nature Photonics}} \textbf{3}, 706--714 (2009).

\bibitem{mottola2023optical}
R.~Mottola, G.~Buser, and P.~Treutlein, \enquote{{Optical memory in a microfabricated rubidium vapor cell},} {\protect\JournalTitle{Physical Review Letters}} \textbf{131}, 260801 (2023).

\bibitem{chang2022frequency}
P.~Chang, H.~Shi, J.~Miao, \emph{et~al.}, \enquote{{Frequency-stabilized Faraday laser with 10$^{-14}$ short-term instability for atomic clocks},} {\protect\JournalTitle{Applied Physics Letters}} \textbf{120}, 141102 (2022).

\bibitem{shi2022frequency}
H.~Shi, P.~Chang, Z.~Wang, \emph{et~al.}, \enquote{{Frequency stabilization of a Cesium Faraday laser with a double-layer vapor cell as frequency reference},} {\protect\JournalTitle{IEEE Photonics Journal}} \textbf{14}, 1--6 (2022).

\bibitem{griffith2010femtotesla}
W.~C. Griffith, S.~Knappe, and J.~Kitching, \enquote{{Femtotesla atomic magnetometry in a microfabricated vapor cell},} {\protect\JournalTitle{Optics Express}} \textbf{18}, 27167--27172 (2010).

\bibitem{sutter2020recording}
J.~U. Sutter, O.~Lewis, C.~Robinson, \emph{et~al.}, \enquote{{Recording the heart beat of cattle using a gradiometer system of optically pumped magnetometers},} {\protect\JournalTitle{Computers and Electronics in Agriculture}} \textbf{177}, 105651 (2020).

\bibitem{staerkind2024high}
H.~St{\ae}rkind, K.~Jensen, J.~H. M{\"u}ller, \emph{et~al.}, \enquote{{High-field optical cesium magnetometer for magnetic resonance imaging},} {\protect\JournalTitle{PRX Quantum}} \textbf{5}, 020320 (2024).

\bibitem{auzinsh2022wide}
M.~Auzinsh, A.~Sargsyan, A.~Tonoyan, \emph{et~al.}, \enquote{{Wide range linear magnetometer based on a sub-microsized K vapor cell},} {\protect\JournalTitle{Applied Optics}} \textbf{61}, 5749--5754 (2022).

\bibitem{dick1991ultrahigh}
D.~Dick and T.~M. Shay, \enquote{{Ultrahigh-noise rejection optical filter},} {\protect\JournalTitle{Opt. Lett.}} \textbf{16}, 867--869 (1991).

\bibitem{yeh1982dispersive}
P.~Yeh, \enquote{{Dispersive magnetooptic filters},} {\protect\JournalTitle{Applied Optics}} \textbf{21}, 2069--2075 (1982).

\bibitem{uhland2023build}
D.~Uhland, H.~Dillmann, Y.~Wang, and I.~Gerhardt, \enquote{{How to build an optical filter with an atomic vapor cell},} {\protect\JournalTitle{New Journal of Physics}} \textbf{25}, 125001 (2023).

\bibitem{downes2023simple}
L.~Downes, \enquote{{Simple Python tools for modelling few-level atom-light interactions},} {\protect\JournalTitle{Journal of Physics B: Atomic, Molecular and Optical Physics}} \textbf{56}, 223001 (2023).

\bibitem{potvliege2025coombe}
R.~M. Potvliege and S.~A. Wrathmall, \enquote{{CoOMBE: A suite of open-source programs for the integration of the optical Bloch equations and Maxwell-Bloch equations},} {\protect\JournalTitle{Computer Physics Communications}} \textbf{306}, 109374 (2025).

\bibitem{ADMweb}
{Simon Rochester}, \enquote{{Atomic Density Matrix},} \url{https://www.rochesterscientific.com/ADM/} (2020). Accessed: 08-12-24.

\bibitem{bala2022comprehensive}
R.~Bala, J.~Ghosh, and V.~Venkataraman, \enquote{{A comprehensive model for Doppler spectra in thermal atomic vapour},} {\protect\JournalTitle{Journal of Physics B: Atomic, Molecular and Optical Physics}} \textbf{55}, 165003 (2022).

\bibitem{sagle1996measurement}
J.~Sagle, R.~K. Namiotka, and J.~Huennekens, \enquote{{Measurement and modelling of intensity dependent absorption and transit relaxation on the cesium D$_{1}$ line},} {\protect\JournalTitle{Journal of Physics B: Atomic, Molecular and Optical Physics}} \textbf{29}, 2629 (1996).

\bibitem{zentile2015elecsus}
M.~A. Zentile, J.~Keaveney, L.~Weller, \emph{et~al.}, \enquote{{ElecSus: A program to calculate the electric susceptibility of an atomic ensemble},} {\protect\JournalTitle{Computer Physics Communications}} \textbf{189}, 162--174 (2015).

\bibitem{keaveney2018elecsus}
J.~Keaveney, C.~S. Adams, and I.~G. Hughes, \enquote{{ElecSus: Extension to arbitrary geometry magneto-optics},} {\protect\JournalTitle{Computer Physics Communications}} \textbf{224}, 311--324 (2018).

\bibitem{adams2018optics}
C.~S. Adams and I.~G. Hughes, \emph{{Optics f2f: from Fourier to Fresnel}} (Oxford University Press, 2018).

\bibitem{siddons2008absolute}
P.~Siddons, C.~S. Adams, C.~Ge, and I.~G. Hughes, \enquote{{Absolute absorption on rubidium D lines: comparison between theory and experiment},} {\protect\JournalTitle{Journal of Physics B: Atomic, Molecular and Optical Physics}} \textbf{41}, 155004 (2008).

\bibitem{zentile2015atomic}
M.~A. Zentile, D.~J. Whiting, J.~Keaveney, \emph{et~al.}, \enquote{{Atomic Faraday filter with equivalent noise bandwidth less than 1 GHz},} {\protect\JournalTitle{Optics Letters}} \textbf{40}, 2000--2003 (2015).

\bibitem{zentile2015optimization}
M.~A. Zentile, J.~Keaveney, R.~S. Mathew, \emph{et~al.}, \enquote{{Optimization of atomic Faraday filters in the presence of homogeneous line broadening},} {\protect\JournalTitle{Journal of Physics B: Atomic, Molecular and Optical Physics}} \textbf{48}, 185001 (2015).

\bibitem{zentile2014hyperfine}
M.~A. Zentile, R.~Andrews, L.~Weller, \emph{et~al.}, \enquote{{The hyperfine Paschen--Back Faraday effect},} {\protect\JournalTitle{Journal of Physics B: Atomic, Molecular and Optical Physics}} \textbf{47}, 075005 (2014).

\bibitem{hanley2015absolute}
R.~K. Hanley, P.~D. Gregory, I.~G. Hughes, and S.~L. Cornish, \enquote{{Absolute absorption on the potassium D lines: theory and experiment},} {\protect\JournalTitle{Journal of Physics B: Atomic, Molecular and Optical Physics}} \textbf{48}, 195004 (2015).

\bibitem{ponciano2020absorption}
F.~S. Ponciano-Ojeda, F.~D. Logue, and I.~G. Hughes, \enquote{{Absorption spectroscopy and Stokes polarimetry in a $^{87}$Rb vapour in the Voigt geometry with a 1.5 T external magnetic field},} {\protect\JournalTitle{Journal of Physics B: Atomic, Molecular and Optical Physics}} \textbf{54}, 015401 (2020).

\bibitem{briscoe2023voigt}
J.~D. Briscoe, F.~D. Logue, D.~Pizzey, \emph{et~al.}, \enquote{{Voigt transmission windows in optically thick atomic vapours: a method to create single-peaked line centre filters},} {\protect\JournalTitle{Journal of Physics B: Atomic, Molecular and Optical Physics}} \textbf{56}, 105403 (2023).

\bibitem{briscoe2024indirect}
J.~D. Briscoe, D.~Pizzey, S.~A. Wrathmall, and I.~G. Hughes, \enquote{{Indirect measurement of atomic magneto-optical rotation via Hilbert transform},} {\protect\JournalTitle{Journal of Physics B: Atomic, Molecular and Optical Physics}} \textbf{57}, 175401 (2024).

\bibitem{pizzey2022laser}
D.~Pizzey, J.~Briscoe, F.~Logue, \emph{et~al.}, \enquote{{Laser spectroscopy of hot atomic vapours: from’scope to theoretical fit},} {\protect\JournalTitle{New Journal of Physics}} \textbf{24}, 125001 (2022).

\bibitem{reed2018low}
D.~Reed, N.~{\v{S}}ibali{\'c}, D.~Whiting, \emph{et~al.}, \enquote{{Low-drift Zeeman shifted atomic frequency reference},} {\protect\JournalTitle{OSA Continuum}} \textbf{1}, 4--12 (2018).

\bibitem{luo2018signal}
B.~Luo, L.~Yin, J.~Xiong, \emph{et~al.}, \enquote{{Signal intensity influences on the atomic Faraday filter},} {\protect\JournalTitle{Optics Letters}} \textbf{43}, 2458--2461 (2018).

\bibitem{xiong2018characteristics}
J.~Xiong, B.~Luo, L.~Yin, \emph{et~al.}, \enquote{{The characteristics of Ar and Cs mixed Faraday optical filter under different signal powers},} {\protect\JournalTitle{IEEE Photonics Technology Letters}} \textbf{30}, 716--719 (2018).

\bibitem{agnew2024practical}
N.~Agnew, G.~Machin, E.~Riis, and A.~S. Arnold, \enquote{{Practical Doppler broadening thermometry},} in \emph{AIP Conference Proceedings,}  vol. 3230 (AIP Publishing, 2024).

\bibitem{foot2005atomic}
C.~J. Foot, \emph{Atomic physics}, vol.~7 (Oxford university press, 2005).

\bibitem{weller2012measuring}
L.~Weller, T.~Dalton, P.~Siddons, \emph{et~al.}, \enquote{{Measuring the Stokes parameters for light transmitted by a high-density rubidium vapour in large magnetic fields},} {\protect\JournalTitle{Journal of Physics B: Atomic, Molecular and Optical Physics}} \textbf{45}, 055001 (2012).

\bibitem{faraday1846experimental}
M.~Faraday, \enquote{{I. Experimental researches in electricity.—Nineteenth series},} {\protect\JournalTitle{Philosophical Transactions of the Royal Society of London}} \textbf{136}, 1--20 (1846).

\bibitem{budker2002resonant}
D.~Budker, W.~Gawlik, D.~Kimball, \emph{et~al.}, \enquote{{Resonant nonlinear magneto-optical effects in atoms},} {\protect\JournalTitle{Reviews of Modern Physics}} \textbf{74}, 1153 (2002).

\bibitem{siddons2009gigahertz}
P.~Siddons, N.~C. Bell, Y.~Cai, \emph{et~al.}, \enquote{{A gigahertz-bandwidth atomic probe based on the slow-light Faraday effect},} {\protect\JournalTitle{Nature Photonics}} \textbf{3}, 225--229 (2009).

\bibitem{aplet1964faraday}
L.~Aplet and J.~W. Carson, \enquote{{A Faraday effect optical isolator},} {\protect\JournalTitle{Applied Optics}} \textbf{3}, 544--545 (1964).

\bibitem{wu1986optical}
Z.~Wu, M.~Kitano, W.~Happer, \emph{et~al.}, \enquote{{Optical determination of alkali metal vapor number density using Faraday rotation},} {\protect\JournalTitle{Applied Optics}} \textbf{25}, 4483--4492 (1986).

\bibitem{portalupi2016simultaneous}
S.~L. Portalupi, M.~Widmann, C.~Nawrath, \emph{et~al.}, \enquote{{Simultaneous Faraday filtering of the Mollow triplet sidebands with the Cs-D1 clock transition},} {\protect\JournalTitle{Nature Communications}} \textbf{7}, 13632 (2016).

\bibitem{mottola2023electromagnetically}
R.~Mottola, G.~Buser, and P.~Treutlein, \enquote{{Electromagnetically induced transparency and optical pumping in the hyperfine Paschen-Back regime},} {\protect\JournalTitle{Physical Review A}} \textbf{108}, 062820 (2023).

\bibitem{muroo1994resonant}
K.~Muroo, T.~Matsunobe, Y.~Shishido, \emph{et~al.}, \enquote{{Resonant Voigt-effect spectrum of the rubidium D$_{2}$ transition},} {\protect\JournalTitle{JOSA B}} \textbf{11}, 409--414 (1994).

\bibitem{kudenov2020dual}
M.~W. Kudenov, B.~Pantalone, and R.~Yang, \enquote{{Dual-beam potassium Voigt filter for atomic line imaging},} {\protect\JournalTitle{Applied Optics}} \textbf{59}, 5282--5289 (2020).

\bibitem{ge2024voigt}
Z.~Ge, C.~Zhu, Y.~Wang, \emph{et~al.}, \enquote{{A Voigt laser lasing on Cs 852 nm transition},} {\protect\JournalTitle{IEEE Access}} \textbf{12}, 196171--196177 (2024).

\bibitem{liu2023atomic}
Z.~Liu, X.~Guan, X.~Qin, \emph{et~al.}, \enquote{{An atomic filter laser with a compact Voigt anomalous dispersion optical filter},} {\protect\JournalTitle{Applied Physics Letters}} \textbf{123}, 131103 (2023).

\bibitem{palik1970infrared}
E.~Palik and J.~Furdyna, \enquote{{Infrared and microwave magnetoplasma effects in semiconductors},} {\protect\JournalTitle{Reports on Progress in Physics}} \textbf{33}, 1193 (1970).

\bibitem{rotondaro2015generalized}
M.~D. Rotondaro, B.~V. Zhdanov, and R.~J. Knize, \enquote{{Generalized treatment of magneto-optical transmission filters},} {\protect\JournalTitle{JOSA B}} \textbf{32}, 2507--2513 (2015).

\bibitem{edwards1995magneto}
N.~Edwards, S.~Phipp, and P.~Baird, \enquote{{Magneto-optic rotation for an arbitrary field direction},} {\protect\JournalTitle{Journal of Physics B: Atomic, Molecular and Optical Physics}} \textbf{28}, 4041 (1995).

\bibitem{nienhuis1998magneto}
G.~Nienhuis and F.~Schuller, \enquote{{Magneto-optical effects of saturating light for arbitrary field direction},} {\protect\JournalTitle{Optics Communications}} \textbf{151}, 40--45 (1998).

\bibitem{keaveney2018optimized}
J.~Keaveney, S.~A. Wrathmall, C.~S. Adams, and I.~G. Hughes, \enquote{{Optimized ultra-narrow atomic bandpass filters via magneto-optic rotation in an unconstrained geometry},} {\protect\JournalTitle{Optics Letters}} \textbf{43}, 4272--4275 (2018).

\bibitem{higgins2020atomic}
C.~R. Higgins, D.~Pizzey, R.~S. Mathew, and I.~G. Hughes, \enquote{{Atomic line versus lens cavity filters: a comparison of their merits},} {\protect\JournalTitle{OSA Continuum}} \textbf{3}, 961--970 (2020).

\bibitem{alqarni2024device}
S.~A. Alqarni, J.~D. Briscoe, C.~R. Higgins, \emph{et~al.}, \enquote{{A device for magnetic-field angle control in magneto-optical filters using a solenoid-permanent magnet pair},} {\protect\JournalTitle{Review of Scientific Instruments}} \textbf{95}, 035103 (2024).

\bibitem{logueThesis}
F.~D. Logue, \enquote{{Improving Magneto-Optical Filter Performance: Cascading and Oblique B-fields.}} Ph.D. thesis, Durham University (2023).

\bibitem{ElecSusGitHub}
{Mark Zentile and James Keaveney}, \enquote{{ElecSus},} \url{https://github.com/durham-qlm/ElecSus} (2018). Accessed: 11-04-25.

\bibitem{wellerThesis}
L.~Weller, \enquote{{Absolute Absorption and Dispersion in a Thermal Rb Vapour at High Densities and High Magnetic Field},} Ph.D. thesis, Durham University (2013).

\bibitem{jones1941new}
R.~C. Jones, \enquote{{A New Calculus for the Treatment of Optical Systems I. Description and Discussion of the Calculus},} {\protect\JournalTitle{J. Opt. Soc. Am.}} \textbf{31}, 488--493 (1941).

\bibitem{griffiths2023introduction}
D.~J. Griffiths, \emph{{Introduction to electrodynamics}} (Cambridge University Press, 2023).

\bibitem{keaveney2018automated}
J.~Keaveney, \enquote{{Automated translating beam profiler for in situ laser beam spot-size and focal position measurements},} {\protect\JournalTitle{Review of Scientific Instruments}} \textbf{89}, 035114 (2018).

\bibitem{sherlock2009weak}
B.~E. Sherlock and I.~G. Hughes, \enquote{{How weak is a weak probe in laser spectroscopy?}} {\protect\JournalTitle{American Journal of Physics}} \textbf{77}, 111--115 (2009).

\bibitem{haupl2025modelling}
D.~R. H{\"a}upl, C.~R. Higgins, D.~Pizzey, \emph{et~al.}, \enquote{{Modelling spectra of hot alkali vapour in the saturation regime},} {\protect\JournalTitle{New Journal of Physics}} \textbf{27}, 033003 (2025).

\bibitem{sargsyan2012hyperfine}
A.~Sargsyan, G.~Hakhumyan, C.~Leroy, \emph{et~al.}, \enquote{{Hyperfine Paschen--Back regime realized in Rb nanocell},} {\protect\JournalTitle{Optics Letters}} \textbf{37}, 1379--1381 (2012).

\bibitem{ciampini2017optical}
D.~Ciampini, R.~Battesti, C.~Rizzo, and E.~Arimondo, \enquote{{Optical spectroscopy of a microsized Rb vapor sample in magnetic fields up to 58 T},} {\protect\JournalTitle{Physical Review A}} \textbf{96}, 052504 (2017).

\bibitem{staerkind2023precision}
H.~St{\ae}rkind, K.~Jensen, J.~H. M{\"u}ller, \emph{et~al.}, \enquote{{Precision measurement of the excited state land{\'e} g-factor and diamagnetic shift of the cesium D$_{2}$ line},} {\protect\JournalTitle{Physical Review X}} \textbf{13}, 021036 (2023).

\bibitem{hughes2010measurements}
I.~Hughes and T.~Hase, \emph{{Measurements and their uncertainties: a practical guide to modern error analysis}} (OUP Oxford, 2010).

\bibitem{rotondaro1997collisional}
M.~D. Rotondaro and G.~P. Perram, \enquote{{Collisional broadening and shift of the rubidium D$_{1}$ and D$_{2}$ lines (52S$_{12}$→ 52P$_{12}$, 52P$_{32}$) by rare gases, H$_{2}$, D$_{2}$, N$_{2}$, CH$_{4}$ and CF$_{4}$},} {\protect\JournalTitle{Journal of Quantitative Spectroscopy and Radiative Transfer}} \textbf{57}, 497--507 (1997).

\bibitem{data}
J.~D. Briscoe, \enquote{{Light propagation through an atomic vapor with non-orthogonal electric field modes [dataset]. Durham University Collections.}} \url{http://doi.org/10.15128/r24m90dv58f} (2025).

\end{thebibliography}


\end{document}


\maketitle

\section{Comparison of ElecSus 3 and ElecSus 4 Over a Larger Parameter Space}

Paper Figs.~3 and 5 not only demonstrated the existence of non-orthogonal electric field modes in a simple atomic vapor system, but also highlighted two of many regimes where significant disagreement between outputs from \emph{ElecSus}~$3$ and \emph{ElecSus}~$4$ can be seen. This disagreement was clearly largest in the regions of strongest mode non-orthogonality. In this section, we explore a larger parameter space in both Regime I and II to highlight that, in general, there is significant disagreement between the predictions of the old and new light propagation formalisms. This further emphasizes that the correct light propagation formalism must account for an atomic vapor system which can exhibit non-orthogonal electric field modes. 

In this section, we chose to investigate atomic spectroscopy in both regimes as a function of $\theta_{B}$ and $\theta_{E}$, since these are the two main parameters which affect light propagation. This can be seen be expanding the complex coefficient equation, Eq.~(11):

\begin{equation} 
\label{eqnSupp: complexCoefficients}
\begin{split}
\begin{pmatrix} c_{1} \\ c_{2} 
\end{pmatrix} & = 
\begin{pmatrix} 1 & \langle\hat{\mathbf{m}}_{1}\vert\hat{\mathbf{m}}_{2}\rangle \\ \langle\hat{\mathbf{m}}_{2}\vert\hat{\mathbf{m}}_{1}\rangle & 1
\end{pmatrix}^{-1}\begin{pmatrix} m_{1, x}^{*} & m_{1, y}^{*} & m_{1, z}^{*} \\ m_{2, x}^{*} & m_{2, y}^{*} & m_{2, z}^{*} 
\end{pmatrix}\begin{pmatrix} \epsilon_{\mathrm{inc}, x} \\ \epsilon_{\mathrm{inc}, y} \\ 0
\end{pmatrix} \,, \\ & = \frac{1}{1 - \langle\hat{\mathbf{m}}_{1}\vert\hat{\mathbf{m}}_{2}\rangle\langle\hat{\mathbf{m}}_{2}\vert\hat{\mathbf{m}}_{1}\rangle} \begin{pmatrix} m_{1, x}^{*}\epsilon_{x} + m_{1, y}^{*}\epsilon_{y} - \langle\hat{\mathbf{m}}_{1}\vert\hat{\mathbf{m}}_{2}\rangle(m_{2, x}^{*}\epsilon_{x} + m_{2, y}^{*}\epsilon_{y}) \\ m_{2, x}^{*}\epsilon_{x} + m_{2, y}^{*}\epsilon_{y} - \langle\hat{\mathbf{m}}_{2}\vert\hat{\mathbf{m}}_{1}\rangle(m_{1, x}^{*}\epsilon_{x} + m_{1, y}^{*}\epsilon_{y})\end{pmatrix} \,,
\end{split}
\end{equation}

\noindent where the incident ``inc'' notation is dropped for clarity, the electric field components $\epsilon_{j}$ depend on $\theta_{E}$, and the modes $\hat{\mathbf{m}}_{i} = [m_{i, x}, m_{i, y}, m_{i, z}]^{\mathrm{T}}$ strongly depend on $\theta_{B}$. While $\hat{\mathbf{m}}_{i}$ also depend on other atomic parameters (via the electric susceptibility components $\chi_{q}$), it is oblique atom-light geometries ($\theta_{B} \neq m\pi/2, m \in \mathbb {Z}$) which cause modes to exhibit non-orthogonality. We therefore gain the most insight into validating the new light propagation formalism by investigating the atom-light interaction as a function of $\theta_{B}$ and $\theta_{E}$, as was done in the paper.

\subsection{Regime I}

\begin{figure}
\centering
{\includegraphics[width=\linewidth]{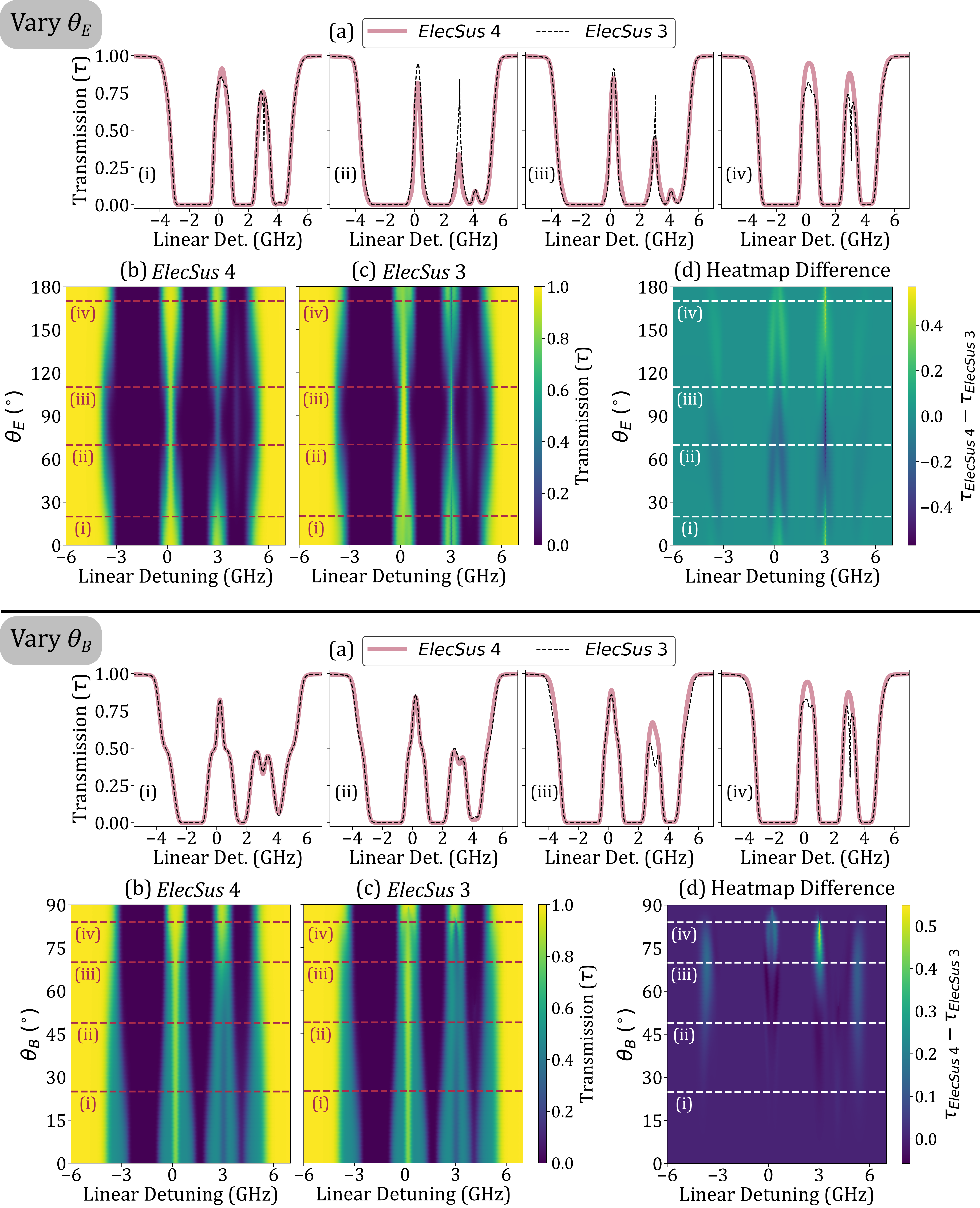}}
\caption{Evolution of the theoretical transmission spectra ($\tau$) of a Rb vapor cell of natural abundance on the D$_{2}$ line. The fixed atomic system parameters are: cell length $L = 2~\mathrm{mm}$, atom temperature $T = 120~^{\circ}\mathrm{C}$, and magnetic field magnitude $B = 240~\mathrm{G}$ ($B = \lvert\mathbf{B}\rvert$). The top sub-figure shows $\tau$ evolution as a function of the linear polarization angle $\theta_{E}$ with respect to the $\hat{\mathbf{x}}$ axis (fixed $\theta_{B} = 84^{\circ}$). The bottom sub-figure shows $\tau$ evolution as a function of the angle $\theta_{B}$ between $\mathbf{B}$ and the light wavevector (fixed $\theta_{E} = 0^{\circ}$). Each sub-figure follows the same structure (top/bottom): in panel (a), the spectra are plotted for several fixed values of $\theta_{E}/\theta_{B}$. These spectra are slices of the heatmaps (b)-(c), highlighted by horizontal red/white dashed lines, and labeled (i)-(iv). Predictions of the theory models \emph{ElecSus}~$4$ (red) and \emph{ElecSus}~$3$ (black, dashed) are shown. The full evolution of $\tau$ over a larger range of $\theta_{E}/\theta_{B}$ is shown for both models in panels (b) and (c); the difference of these heatmaps is plotted in panel (d).}
\label{fig: transmissionEvolutionRegimeI}
\end{figure}

We start with Regime I: in Fig.~\ref{fig: transmissionEvolutionRegimeI}, we plot theoretical transmission spectra $\tau$ modeling a weak-probe laser beam traversing a naturally abundant Rb vapor cell. Each spectrum is plotted as a function of linear detuning $\Delta$ from a global line-center for an atomic system that has the following parameters: D-line = D$_{2}$, $L = 2~\mathrm{mm}$, $T = 120~^{\circ}\mathrm{C}$, $B = 240~\mathrm{G}$, and $\Gamma_{\mathrm{B}} = 0~\mathrm{MHz}$. The figure is separated into two sub-figures; the top sub-figure shows spectra evolution as a function of $\theta_{E}$ (fixed $\theta_{B} = 84^{\circ}$), while the bottom sub-figure shows spectra evolution as a function of $\theta_{B}$ (fixed $\theta_{E} = 0^{\circ}$). Both sub-figures follow exactly the same structure. The heatmaps use color to show $\tau$ evolution as a function of $\theta_{E}$ or $\theta_{B}$: (b) uses \emph{ElecSus}~$4$, while (c) uses \emph{ElecSus}~$3$. The heatmap (d) shows the difference of these heatmaps i.~e.~(d) = (b) - (c). In panel (a), we plot 4 slices of the heatmaps (b) and (c). For the top sub-figure, $\theta_{E}$ = : (i) $20^{\circ}$, (ii) $70^{\circ}$, (iii) $110^{\circ}$, and (iv) $170^{\circ}$. For the bottom sub-figure, $\theta_{B}$ = : (i) $25^{\circ}$, (ii) $49^{\circ}$, (iii) $70^{\circ}$, and (iv) $84^{\circ}$. These slices are indicated on each heatmap by a red or white dashed line. 

For magnetic field angle variation, the range $\theta_{B} = 0^{\circ}-90^{\circ}$ shows the transition between the Faraday ($\theta_{B} = m\pi, m \in \mathbb {Z}$) and Voigt geometries ($\theta_{B} = (2m+1)\pi/2, m \in \mathbb {Z}$) respectively. Since these geometries have orthogonal electric field modes, we expect (and find) complete agreement between the old and new light propagation formalisms for $\theta_{B} = 0^{\circ}$ and $\theta_{B} = 90^{\circ}$. The heatmaps show regions inbetween these geometries where clear deviation between \emph{ElecSus}~$3$ and \emph{ElecSus}~$4$ outputs can be seen, particularly when $\theta_{B}$ is close to $\theta_{B} = 90^{\circ}$. It is also clear that for this magnetic field magnitude ($B = 240~\mathrm{G}$), there are large frequency ranges where both models agree well. As discussed in the paper, any deviation is caused by a non-zero mode overlap. In general, the stronger the mode overlap, the larger the disagreement between \emph{ElecSus}~$3$ and \emph{ElecSus}~$4$ outputs. The largest transmission disagreement shown in the $\theta_{B}$ heatmap difference plot is 0.55 ($55\%$ light transmission) for the angle $\theta_{B} \approx 84^{\circ}$. As shown in panel (a)(iv) of the bottom sub-figure in Fig.~\ref{fig: transmissionEvolutionRegimeI} ($\theta_{B} = 84^{\circ}$), the disagreement is large because \emph{ElecSus}~$3$ predicts a sharp unphysical feature at $\Delta \approx 3~\mathrm{GHz}$ where the modes significantly overlap. This feature is completely removed by the new light propagation formalism integrated into \emph{ElecSus}~$4$.

Large mode overlap doesn't necessarily imply significant disagreement between the old and new light propagation formalisms. As shown by Eq.~(\ref{eqnSupp: complexCoefficients}), the complex coefficients (which account for mode non-orthogonality) also depend on the incident electric field polarization, and therefore $\theta_{E}$. This is the mathematical representation of the atom-light coupling. We therefore expect light polarization to modify the disagreement between \emph{ElecSus}~$3$ and \emph{ElecSus}~$4$ outputs, even for fixed atomic parameters. The top sub-figure of Fig.~\ref{fig: transmissionEvolutionRegimeI} shows $\tau$ evolution over the range $\theta_{E} = 0^{\circ}-180^{\circ}$ for 
fixed $\theta_{B} = 84^{\circ}$, where we previously saw significant disagreement between models (for $\theta_{E} = 0^{\circ}$). The heatmaps again show large frequency ranges where both models agree well. As expected, significant disagreement is seen where the modes show the strongest overlap, but the degree of disagreement depends on $\theta_{E}$. The largest positive transmission disagreement shown in the $\theta_{E}$ heatmap difference plot is 0.57 ($57\%$ light transmission) for $\theta_{E} \approx 172^{\circ}$, while the largest negative transmission disagreement shown is -0.57 for $\theta_{E} \approx 82^{\circ}$. This highlights that \emph{ElecSus}~$3$ can both under and over-predict light transmission where modes have a non-zero overlap. It is very clear from Fig.~\ref{fig: transmissionEvolutionRegimeI} that the new light propagation formalism corrects the outputs of the old formalism in Regime I over a much larger parameter space than what could feasibly be presented in the paper.

\subsection{Regime II}

\begin{figure}
\centering
{\includegraphics[width=\linewidth]{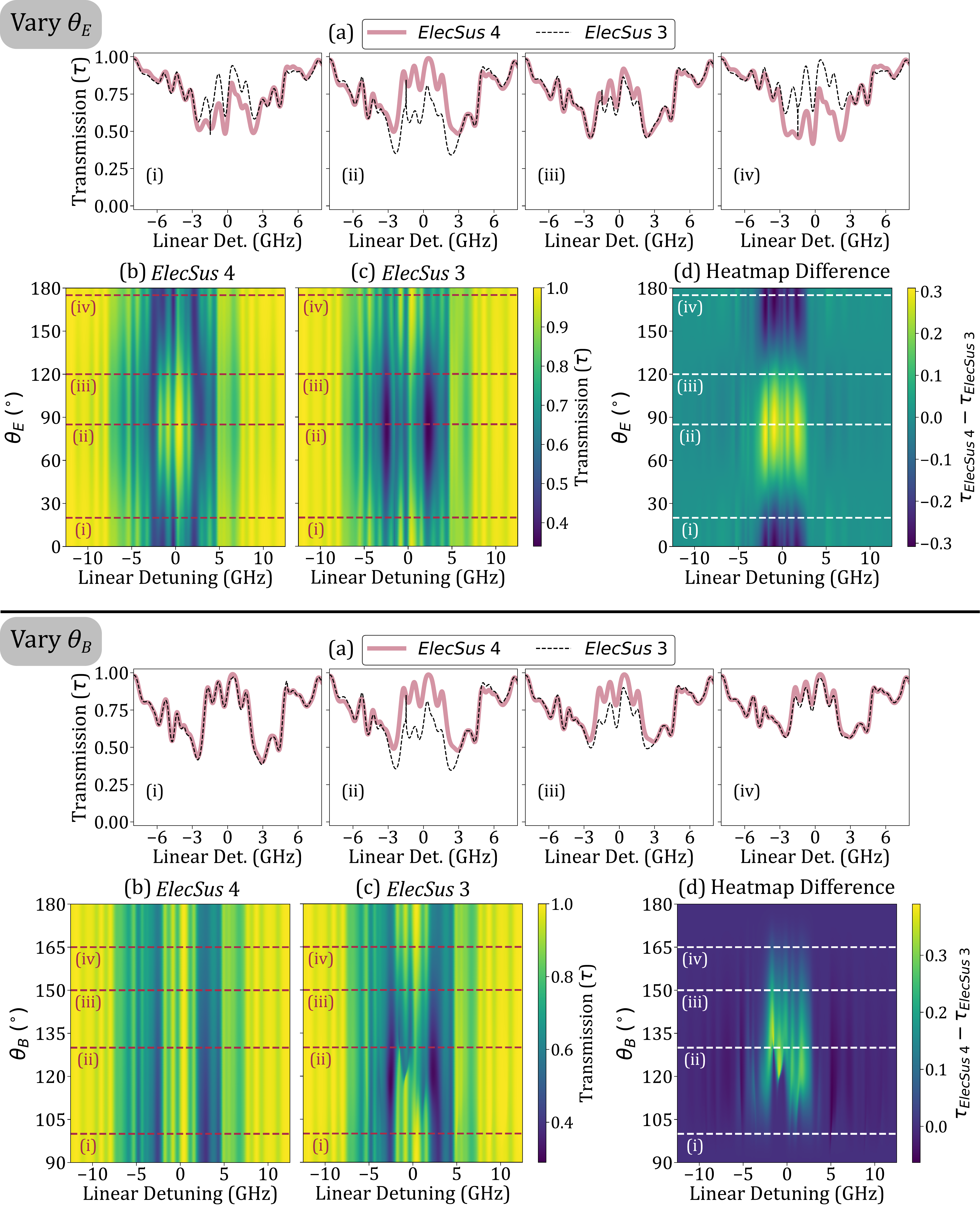}}
\caption{Evolution of the theoretical transmission spectra ($\tau$) of a Rb vapor cell of natural abundance on the D$_{2}$ line. The fixed atomic system parameters are: cell length $L = 2~\mathrm{mm}$, atom temperature $T = 80~^{\circ}\mathrm{C}$, and magnetic field magnitude $B = 2.5~\mathrm{kG}$ ($B = \lvert\mathbf{B}\rvert$). The top sub-figure shows $\tau$ evolution as a function of the linear polarization angle $\theta_{E}$ with respect to the $\hat{\mathbf{x}}$ axis (fixed $\theta_{B} = 130^{\circ}$). The bottom sub-figure shows $\tau$ evolution as a function of the angle $\theta_{B}$ between $\mathbf{B}$ and the light wavevector (fixed $\theta_{E} = 90^{\circ}$). Each sub-figure follows the same structure (top/bottom): in panel (a), the spectra are plotted for several fixed values of $\theta_{E}/\theta_{B}$. These spectra are slices of the heatmaps (b)-(c), highlighted by horizontal red/white dashed lines, and labeled (i)-(iv). Predictions of the theory models \emph{ElecSus}~$4$ (red) and \emph{ElecSus}~$3$ (black, dashed) are shown. The full evolution of $\tau$ over a larger range of $\theta_{E}/\theta_{B}$ is shown for both models in panels (b) and (c); the difference of these heatmaps is plotted in panel (d).}
\label{fig: transmissionEvolutionRegimeII}
\end{figure}

Fig.~\ref{fig: transmissionEvolutionRegimeII} is the Regime II equivalent of Fig.~\ref{fig: transmissionEvolutionRegimeI}; we therefore only discuss key changes and observations in this section. The following changes are made to the atomic system global parameters: $T = 80~^{\circ}\mathrm{C}$, and $B = 2.5~\mathrm{kG}$. The top sub-figure shows spectra evolution as a function of $\theta_{E}$ (fixed $\theta_{B} = 130^{\circ}$), while the bottom sub-figure shows spectra evolution as a function of $\theta_{B}$ (fixed $\theta_{E} = 90^{\circ}$). For the top sub-figure, $\theta_{E}$ = : (i) $20^{\circ}$, (ii) $85^{\circ}$, (iii) $120^{\circ}$, and (iv) $175^{\circ}$. For the bottom sub-figure, $\theta_{B}$ = : (i) $100^{\circ}$, (ii) $130^{\circ}$, (iii) $150^{\circ}$, and (iv) $165^{\circ}$. 

For magnetic field angle variation, the transition between the Faraday and Voigt geometries is now $\theta_{B} = 90^{\circ}-180^{\circ}$ respectively, to explore a different quadrant of $\theta_{B}$ parameter space. We again find complete agreement between the old and new light propagation formalisms in these geometries, as expected. The deviation between \emph{ElecSus}~$3$ and \emph{ElecSus}~$4$ outputs is even clearer than in Regime I, shown by a much larger area of non-zero transmission difference on the heatmap difference plot. While the maximum difference magnitude is not as large (0.39, or $39\%$, for $\theta_{B} = 130^{\circ}$), the disagreement between models covers a much larger range in linear detuning when compared to Regime I. We believe this is because the magnetic field magnitude in Regime II is an order of magnitude larger than in Regime I; the Zeeman shift frequency shifts atomic resonances in both the red and blue detuning directions, increasing the region of mode non-orthogonality in frequency space. As shown in panel (a)(ii) of the bottom sub-figure in Fig.~\ref{fig: transmissionEvolutionRegimeII} ($\theta_{B} = 130^{\circ}$), the disagreement is again large because \emph{ElecSus}~$3$ predicts a sharp unphysical feature at $\Delta \approx -1.5~\mathrm{GHz}$ where the modes significantly overlap. The new light propagation formalism integrated into \emph{ElecSus}~$4$ completely removes this feature, as it did in Regime I.

For linear polarization angle variation, the top sub-figure of Fig.~\ref{fig: transmissionEvolutionRegimeII} shows $\tau$ evolution over the range $\theta_{E} = 0^{\circ}-180^{\circ}$ for fixed $\theta_{B} = 130^{\circ}$. We use the same $\theta_{E}$ range, since the output spectra is invariant under the permutation $\hat{\boldsymbol{\epsilon}}_{\mathrm{inc, lin}} \rightarrow -\hat{\boldsymbol{\epsilon}}_{\mathrm{inc, lin}}$ for an incident linearly polarized electric field. The transmission difference evolution behaves identically to Regime I, varying between a large positive transmission disagreement and a large negative transmission disagreement. As for $\theta_{B}$ variation, the frequency range of disagreement is much larger when compared to Regime I. It is very clear from Fig.~\ref{fig: transmissionEvolutionRegimeII} that the new light propagation formalism corrects the outputs of the old formalism over a very large region of parameter space.

\section{Fitting with ElecSus 3 (Old Light Propagation Formalism)}
\label{secSupp: ElecSus3Fits}

\begin{figure}
\centering
{\includegraphics[width=\linewidth]{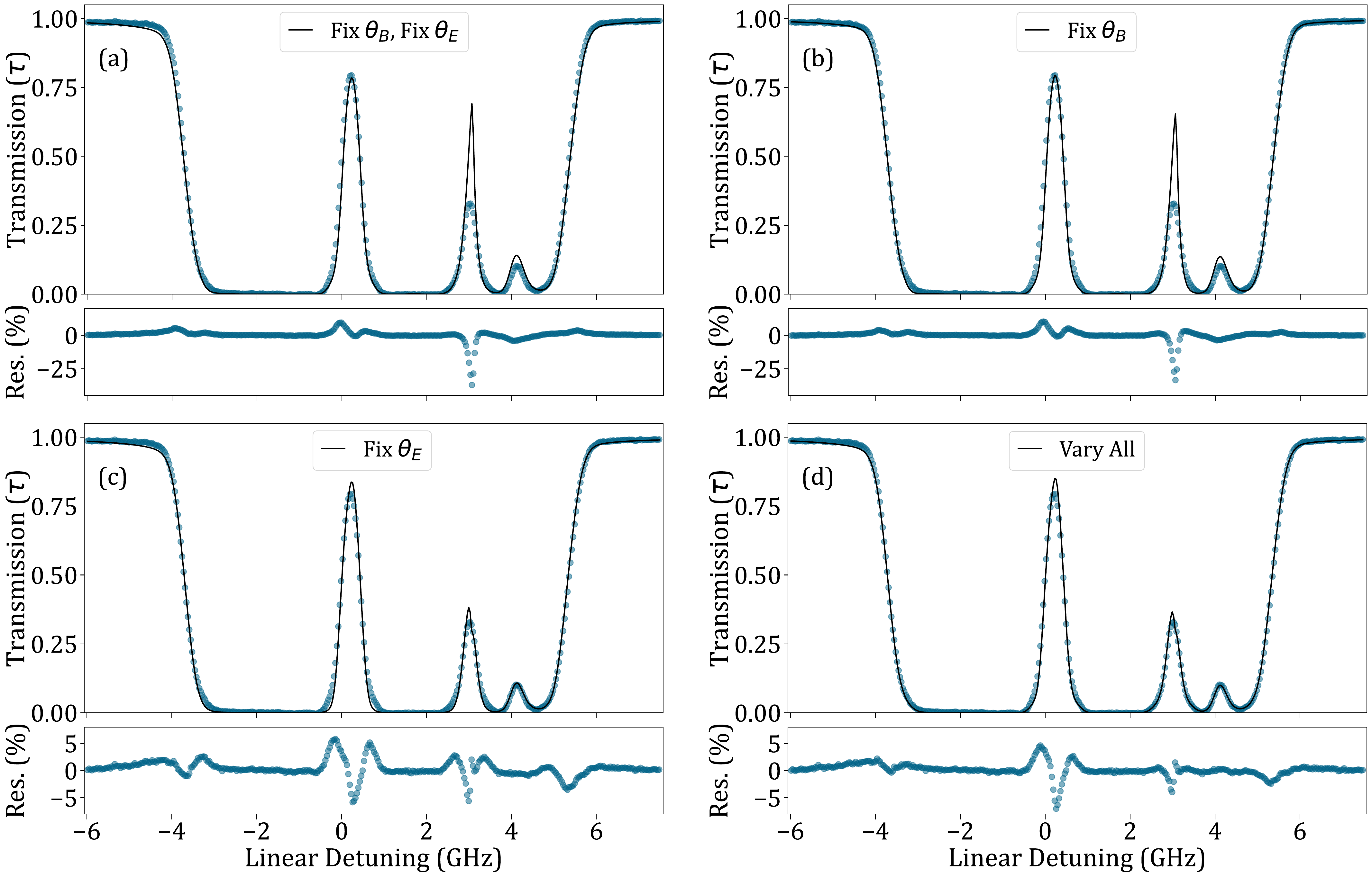}}
\centering
\begin{tabular}{ccccccc}
\hline
Panel & Type& $T$ & $B$ & $\theta_{B}$ & $\theta_{E}$ & $\Gamma_{\mathrm{B}}$ \\
& & ($^{\circ}\mathrm{C}$) & (G) & ($^{\circ}$) & ($^{\circ}$) & $(\mathrm{MHz})$ \\
\hline
(a) & Fit & $119.66\pm0.01$ & \underline{$258.4\pm0.01$} & $80$ & $90$ & $24.7\pm0.1$ \\
 & Bounds & [$110.6$, $135.2$] & [$211.5$, $258.5$] & - & - & [$0$, $30$] \\
(b) & Fit & $119.65\pm0.01$ & \underline{$258.2\pm0.02$} & $80$ & \underline{$100$} & $16.5\pm0.1$ \\
& Bounds & [$110.6$, $135.2$] & [$211.5$, $258.5$] & - & [$80$, $100$] & [$0$, $30$] \\
(c) & Fit & $118.94\pm0.01$ & $236.60\pm0.05$ & $85.65\pm0.01$ & $90$ & $25.1\pm0.1$ \\
& Bounds & [$110.6$, $135.2$] & [$211.5$, $258.5$] & [$70$, $90$] & - & [$0$, $30$] \\
(d) & Fit & $119.83\pm0.01$ & $238.65\pm0.04$ & $86.04\pm0.01$ & \underline{$80.02\pm0.01$} & $20.9\pm0.1$ \\
& Bounds & [$110.6$, $135.2$] & [$211.5$, $258.5$] & [$70$, $90$] & [$80$, $100$] & [$0$, $30$] \\
\hline
\end{tabular}
\caption{Experimental transmission spectroscopy of a natural abundance Rb vapor cell of length $L = 2~\mathrm{mm}$. The atoms are subject to an external magnetic field ($B = \vert\mathbf{B}\vert$) oriented at $\theta_{B} = 80^{\circ}$ relative to the light wavevector, and incident linear polarization angle $\theta_{E} \approx 90^{\circ}$ (relative to $\hat{\mathbf{x}}$). Data (blue) is compared to \emph{ElecSus}~$3$ fits (black) using different fit parameter bounds and fixed parameters. Extracted fit parameters and bounds are indicated in the table, attached to a panel/spectrum (a)-(d). Lower and upper bounds are indicated using square brackets. These were chosen based on measurements/calibrations from the experiment. A fixed parameter (with no bound) is set to that value in the fit; fixed parameters are also indicated on each panel. Parameters which fit to boundary values are underlined for clarity.}
\label{fig: elecSus3Fits}
\end{figure}

In the results section of the paper, we compare data taken in Regimes I and II with \emph{ElecSus}~$4$ fits. \emph{ElecSus}~$3$ fits were omitted, primarily due to fitting to parameters which do not match those measured in the lab. This is not a surprising result, given that one of the main aims of the paper is to correct the old light propagation formalism which is integrated into \emph{ElecSus}~$3$. Parameters of a fit can be fixed to ensure that the most important parameters, such as $\theta_{B}$, match those measured in the lab. However, as shown by paper Figs.~3 and 5, \emph{ElecSus}~$3$ outputs have unphysical features which can contribute large residuals to a fit. This is ultimately why the fit algorithm deviates from measured parameters. In this section, we show an example of \emph{ElecSus}~$3$ fitting using the $\theta_{E} \approx 90^{\circ}$ dataset in Regime I (see paper Fig.~4a). Using the same figure, this dataset is an excellent choice, shown by the large unphysical feature in the \emph{ElecSus}~$3$ output (using \emph{ElecSus}~$4$ fit parameters). 

The results are shown in Fig.~\ref{fig: elecSus3Fits}. We take an average of the 4 repeats taken for the $\theta_{E} \approx 90^{\circ}$ dataset, which is plotted in blue on each of the sub-panels (a)-(d). Each panel shows an \emph{ElecSus}~$3$ fit with different fit constraints. These are: (a) fix $\theta_{B} = 80^{\circ}$ and $\theta_{E} = 90^{\circ}$, (b) fix $\theta_{B} = 80^{\circ}$, (c) fix $\theta_{E} = 90^{\circ}$, and (d) vary all fit parameters. For all panels, $T$ and $B$ are allowed to float within bounds set to $\pm10\%$ of the measured value. If $\theta_{B}$ or $\theta_{E}$ can float, the bounds are set to $\pm10^{\circ}$ of the measured value (the value when fixed). We expect $\Gamma_{\mathrm{B}} = 5-10~\mathrm{MHz}$ based on fitting to high temperature, zero $B$ field spectra; a conservative bound of $0-30~\mathrm{MHz}$ is therefore set. These bounds are shown in the table positioned above the figure caption. We choose sensible bounds that avoid the possibility of the algorithm missing a true global minimum. The bounds are large enough that it is easy to conclude if the fit parameters are not within experiment measurement error. Having set the constraints and bounds, the fit is run 100 times for each set of constraints to ensure a global minimum is found. Due to complex nature of the parameter space, we choose a differential evolution fit algorithm as we typically find this is best for avoiding local minima. The reduced mean squared error (RMSE) of each fit is calculated, and used to filter the 100 fits for those with the lowest RMSE (and therefore the best fit). We then take the mean and standard error for each fit parameter~\cite{hughes2010measurements}; the results are shown in the figure table. Due to the large number of runs of the fit, the fit errors are tiny, representing a small spread of the fit parameter and a large number of iterations of the fit. We underline any parameters which can vary, but return a value which lies on the bounds of that parameter. This implies that if the bound was larger (and therefore further from the measured value), the fit would be better. Residuals are plotted in Fig.~\ref{fig: elecSus3Fits}, and in all cases are worse than the \emph{ElecSus}~$4$ fits in the paper. We clearly see that by applying constraints (fixing values to match lab measurements), parameters such as $\theta_{B}$ cause large unphysical features with large residuals to the data. For both (a) and (b), the lineshape is clearly incorrect, and the fit parameters---e.~g.~$B$ for (a) and $\theta_{E}$ for (b)---return a boundary value, far from the measured value in the lab (at the center of the bounds). Lifting the $\theta_{B}$ constraint in (c) and (d) improves the lineshape, but $\theta_{B}$ no longer matches the measured lab value. In both cases, $\theta_{B}$ is $6^{\circ}$ away from the measured value; this far exceeds the cautious $\alpha_{\theta_{B}} = 0.5^{\circ}$ error set by using half an increment on the large rotation platform that rotated the magnets (see paper Fig.~2). It is clear that \emph{ElecSus}~$3$ fits want to avoid the unphysical features caused by non-orthogonal modes, which the new light propagation formalism implemented into \emph{ElecSus}~$4$ clearly accounted for.

\bibliography{bibliography}